 \documentclass[a4paper]{elsarticle}

\usepackage{amssymb}
\usepackage{lipsum}
\usepackage{xcolor}
\usepackage{cancel}
\usepackage{amsmath}

\journal{Annals of Physics}

\begin{document}

\begin{frontmatter}



\title{Hartman Effect from a Geometrodynamic Extension of Bohmian Mechanics}


\author[first,second]{Said Lantigua}
\affiliation[first]{organization={National Laboratory for Scientific Computing},
            addressline={Getulio Vargas Avenue, 333 - Quitandinha}, 
            city={Petropolis},
            postcode={25651-075}, 
            state={RJ},
            country={Brazil}}
\author[second]{Jonas Maziero}
\affiliation[second]{organization={Departament of Physics, Center for Natural and Exact Sciences, Federal University of Santa Maria},
            addressline={Building 13, Roraima Avenue, 1000 - Camobi}, 
            city={Santa Maria},
            postcode={97105-900}, 
            state={RS},
            country={Brazil}}

\begin{abstract}
This paper develops a geometrodynamic extension of Bohmian mechanics to describe quantum tunneling through a potential barrier, treating particle trajectories as geodesics in an Alcubierre-type spacetime. The model provides analytical expressions for the quantum potential, particle dynamics, and tunneling time, explicitly linked to the underlying spacetime geometry. For narrow barriers, the tunneling time depends on the barrier width, while for sufficiently wide barriers, it saturates to a constant value—recovering the Hartman effect. This behavior arises from a geometric self-regulation mechanism, where the quantum potential dynamically adjusts the spacetime distortion to maintain a fixed tunneling time, consistent with relativistic causality despite effective superluminal propagation. The results establish a direct connection between quantum tunneling and spacetime geometry, offering a unified framework to interpret the Hartman effect. This approach naturally incorporates relativistic constraints while suggesting that similar geometric mechanisms may underlie other quantum phenomena, such as topological phases in condensed matter systems.
\end{abstract}




\end{frontmatter}




\section{Introduction}
\label{sec:1}
The unusual mathematical expression of a tunneling time independent of the barrier width, discovered in Ref. \cite{Hartman196200} and exhibited by particles incident on a very wide potential barrier \cite{Sakurai201100, Zettili200900, Landauer198900}, led to discussions about the possibility of superluminal tunneling velocities \cite{Longhi200100,Longhi202200, Sokolovski200000, Winful200200, Winful200300}. Initially, this triggered a vigorous debate about whether this result, later termed the Hartman effect, was due to inaccuracies in the tunneling time measurements  \cite{Ramos202000,Kiukas200900}. It subsequently forced numerous theorists to address the fundamental question    \cite{Buttiker198300, Hauge198700, Winful200301, Winful200400, Winful200600, Rivlin202100, Rivlin202101, Petersen201700}: What is the correct definition of tunneling time?

Recent experimental work in Ref. \cite{Sharoglazova202500} has reignited this debate by challenging the predictions of Bohmian mechanics regarding dwell times for particles tunneling into infinitely long barriers. Their findings, which show finite dwell times contrary to Bohmian predictions, highlight the ongoing complexity of this question. Similarly, Ref. \cite{Fedrizzi202500} discussed how these measurements add empirical data to a discourse previously dominated by theory, yet the debate remains unresolved due to the lack of definitive predictions from standard quantum mechanics.

This question, captivating not only for physicists in the past but also relevant to contemporary researchers, has endured through the years, with many attempts to answer it proving unsuccessful across various approaches \cite{Winful200400, Winful200600, Martinez200600, Lantigua202300}. However, despite unsuccessful attempts, the Hartman effect has led to interesting applications, such as proposals to exploit this effect for sending signals (information) faster than in free evolution, though this is limited by an apparent trade-off between speedup and intensity due to exponential damping during tunneling \cite{Kiukas200900}.

Despite the considerable body of literature addressing this paradox, existing work primarily focuses on definitions derived from non-relativistic and relativistic quantum mechanics. As mentioned above, these approaches do not provide a satisfactory explanation for the observed effect \cite{Hauge198700,Winful200301,Winful200400,Winful200600,Martinez200600,Lantigua202300,  Bandopadhyay200400,Bandopadhyay202101, Delgado200300, Goldberg196700, Hasan202000, Hasan202100,    Leavens198900,Leavens199500, Muga199200}. In contrast, within the context of Bohmian mechanics, the question of defining tunneling time has not been sufficiently addressed \cite{Hagmann199300, Norsen201300}, despite a considerable body of literature on the application of Bohmian theory to particle scattering over a potential barrier \cite{Norsen201300}. One of the more intriguing works in this context investigates Klein tunneling times for electrons in a two-terminal graphene device comprising a potential barrier between metallic contacts \cite{Winful200600, Pandey201900}.

On the other hand, due to the mathematical structure of Bohmian mechanics \cite{Barut198800, Durr200400,Dewdney198700, Sole201600}, it is possible, through certain physical and mathematical considerations, to extend it to a version that incorporates elements of spacetime as defined in general relativity \cite{Gron200700, Janssen201300}. Therefore, it becomes feasible to develop a geometrodynamic description of quantum systems and to derive mathematical expressions not provided by orthodox relativistic or non-relativistic quantum theory \cite{Greiner200000,Gomez201900}. Consequently, this work presents the construction of a solution that explains the Hartman effect through a geometrodynamic approach to Bohmian mechanics, considering a spacetime endowed with a metric that allows for superluminal travel without violating the principles of general relativity \cite{Einstein192800,Alcubierre199400}.

More specifically, the present work reports a general solution constructed under the hypothesis that particles, during the tunneling phenomenon, follow geodesic trajectories within an Alcubierre-like spacetime \cite{Alcubierre199400, Santos202000}. From this solution, quantities of interest are determined, including mathematical expressions describing linear momenta, quantum potentials, and particle positions for the problem under investigation. As a result, the particle's position expression leads to the mathematical expression of the tunneling time, equipped with the geometric elements of the considered spacetime.

The remainder of this article is organized as follows. Section \ref{sec:2} provides a brief yet necessary review of the fundamental concepts of Bohmian mechanics. Section \ref{sec:3} deduces the geometrodynamic version of the quantum force, starting from some results from the literature \cite{Gomez201900}. Section \ref{sec:4} presents the construction of the general solution to the problem of a particle incident from the left on a barrier of constant potential for each of the regions of interest. In Section \ref{sec:5}, the general expression of the quantum potential for each region of interest is determined. Section \ref{sec:6} establishes the general expression for linear momentum and particle position, again for each region of interest, consequently providing the mathematical expression for the tunneling time. Section \ref{sec:7} presents the conclusion of the work. We also include an Appendix containing some mathematical results and methods used in the main text.

\section{Fundamentals of Bohmian mechanics}
\label{sec:2}
Bohmian mechanics is a theory discovered by Louis de Broglie in 1927 and later rediscovered by David Bohm in 1952. It provides a description of quantum systems through a mathematical object known as the wave function, which encodes their dynamics  \cite{Dewdney198700}. Essentially, the partial description of the quantum system is given by a wave function that evolves according to the Schr\"odinger's equation.

On the other hand, the evolution of particles is described by the guidance equation, which also allows expressing their velocities in terms of this wave function  \cite{Barut198800, Oriols201900}. In summary, Bohmian mechanics is an interpretation of quantum theory in which the physical consequences of the quantum postulates, of the collapse of the wave function, and the hypothesis of quantum equilibrium emerge from an analysis of the equations of motion, namely: the Schr\"odinger and guidance equations \cite{Durr200400}.

In presenting the foundational mathematical framework of Bohmian mechanics, we consider the following wave function:
\begin{equation}\label{eq:1}
    \Psi = \sqrt{\rho} \exp{\{ iS \}},
\end{equation}
where $\rho \equiv \rho (\vec{x},t)$ and $S \equiv S(\vec{x},t)$. Upon substituting this function into the Schr\"odinger's  equation,
\begin{equation}\label{eq:2}
    \biggl[- \frac{\hbar^{2}}{2m} \nabla^{2} + V \biggr]\Psi = i \hbar \partial_{t}\Psi,
\end{equation}
we obtain the expressions:
\begin{equation}\label{eq:3}
    - \frac{\hbar^{2}}{2m} \frac{\nabla^{2} \sqrt{\rho}}{\sqrt{\rho}} + \frac{\hbar^{2}}{2m} (\nabla S)^{2} + V = - \hbar \partial_{t} S
\end{equation}
and
\begin{equation}\label{eq:4}
    - \frac{\hbar^{2}}{2m} \biggl[ 2(\nabla \sqrt{\rho}) \cdot \nabla S + \sqrt{\rho}(\nabla^{2} S) \biggr] = \hbar \partial_{t} \sqrt{\rho}.
\end{equation}
Furthermore, using the equalities $2\sqrt{\rho}(\nabla \sqrt{\rho}) = \nabla \rho$ and $2\sqrt{\rho}(\partial_{t} \sqrt{\rho}) = \partial_{t} \rho$\footnote{We emphasize that the following notation is used: $\partial_{t} () = \frac{\partial()}{\partial t}$ stands for the partial derivative with respect to time.}, we can multiply Eq. (\ref{eq:4}) by $2\sqrt{\rho}$ to obtain:
\begin{equation}\label{eq:5}
    \partial_{t} \rho + \frac{\hbar}{m} \nabla \cdot (\rho \nabla S) = 0.
\end{equation}
Additionally, recalling that the probability current density is defined as $\vec{j} \equiv \rho \vec{v} = \Psi^{\ast} \Psi \vec{v}$, we can determine the expression for particle velocity by substituting Eq. (\ref{eq:1}) into:
\begin{equation}\label{eq:6}
    \Psi^{\ast} \Psi \vec{v} = \frac{i \hbar}{2m} \biggl[ \Psi \nabla \Psi^{\ast} - \Psi^{\ast} \nabla \Psi \biggr],
\end{equation}
resulting in:
\begin{equation}\label{eq:7}
    \vec{v} = \frac{\hbar}{m} \nabla S.
\end{equation}
This allows us to rewrite equations (\ref{eq:3}) and (\ref{eq:5}) as:
\begin{equation}\label{eq:8}
    \hbar \partial_{t} S + \frac{\hbar^{2}}{2m} (\nabla S)^{2} + V + Q = 0
\end{equation}
and
\begin{equation}\label{eq:9}
    \partial_{t} \rho + \nabla \cdot \vec{j} = 0,
\end{equation}
where we introduced the quantity 
\begin{equation}\label{eq:10}
    Q \equiv - \hbar^{2} (\nabla^{2} \sqrt{\rho}) / (2m \sqrt{\rho}),
\end{equation}
referred to as the quantum potential or Bohmian potential, and $\vec{j} \equiv \rho  [\hbar (\nabla S) / m]$. In other words, equations (\ref{eq:8}) and (\ref{eq:9}) are the differential equations describing the behavior of $S$ and $\rho$ \cite{Pena200600}. Finally, to complete the formalism, the gradient of Eq. (\ref{eq:8}) is calculated, and the result is expressed in terms of the velocity field (\ref{eq:7}) to obtain:
\begin{equation}\label{eq:11}
    m \partial_{t} \vec{v} + m(\vec{v} \cdot \nabla) \vec{v} = - \nabla (V + Q) = \vec{F}_{V} + \vec{F}_{Q}.
\end{equation}
The above expression is the equation of motion that a particle with a probability current density given by Eq. (\ref{eq:6}) will follow under the action of an effective force $\vec{F} = \vec{F}_{V} + \vec{F}_{Q}$. Therefore, in the classical limit ($\hbar \rightarrow 0$), the trajectories will obey the laws of Newtonian motion, as expected \cite{Barut198800,Dewdney198700,Oriols201900, Pena200600}.

\section{Geometrodynamics and Quantum Force}
\label{sec:3}
When considering the extension of Bohmian mechanics to a relativistic version, several seemingly insurmountable challenges arise. For instance, issues include extending Bohmian trajectories to relativistic paths, constructing a four-vector of probability density and current, and dealing with non-physical trajectory occurrences—especially concerning photons—due to frames of reference where velocity is zero \cite{Foo202200,Foo202300}. However, through careful physical and mathematical considerations, it is possible to construct geometrodynamic models that allow extending the Bohmian formalism to a non-relativistic version that incorporates space-time elements defined in general relativity \cite{Gron200700, Janssen201300}.

For that reason, the geometrodynamic description of quantum systems allows us to derive mathematical expressions that are not provided by approaches based on the orthodox formalism of relativistic or non-relativistic quantum theory \cite{Gomez201900}. Therefore, this section presents a deduction of the quantum force equation in curvilinear coordinates. In other words, it derives a generalized version of the quantum force expression presented in the literature \cite{Gomez201900}.

In this sense, we would like to highlight that the procedure followed here is essentially the one proposed in the literature \cite{Gomez201900}, and their contributions play a central role in the construction of the solution presented in this article. Hence, a possible extension of the Bohmian formalism that incorporates these elements of general relativity arises by postulating the hypothesis that particles in the quantum system follow geodesic trajectories in spacetime.

Consequently, this hypothesis can be expressed mathematically by the following expression:
\begin{equation}\label{eq:12}
    0 = d_{t}^{2}x^{\mu} + \Gamma^{\mu}_{\alpha \beta}d_{t}x^{\alpha} d_{t}x^{\beta},
\end{equation}
where $\Gamma_{\alpha \beta}^{\mu}$ are the Christoffel symbols, the indices $\mu$, $\alpha$, and $\beta$ take values in $\{0,1,2,3\}$. The hypothesis of geodesic trajectories arises naturally when coupling quantum mechanics to general relativity, where particles follow extreme paths in a curved spacetime. This approach reconciles the principle of least action from classical mechanics with Bohmian quantum dynamics, extending it to relativistic scenarios. In this framework, equation (\ref{eq:12}) yields the geometrodynamic constraint:
\begin{equation}\label{eq:13}
    \Gamma^{0}_{\alpha \beta}d_{t}x^{\alpha} d_{t}x^{\beta} = 0.
\end{equation}
This establishes a local equivalence relation between the Euclidean spacetime where Bohmian mechanics is defined and the spacetime of the Lorentzian manifold by considering $x^{0} = ct$  \cite{Gomez201900}. Therefore, by considering $j,k,\mu = i = 1,2,3$ and $\alpha,\beta = 0,1,2,3$, equation (\ref{eq:12}) can be developed to obtain:
\begin{equation}\label{eq:14}
    0 = d_{t}^{2}x^{i} + \Gamma^{i}_{j k}d_{t}x^{j} d_{t}x^{k} - 2c\Gamma^{i}_{0j}d_{t}x^{j} + c^{2}\Gamma^{i}_{00}.
\end{equation}
Considering the components of the gradient\footnote{Analogously to what was clarified in section \ref{sec:2}, we clarify that, in this article, the following notation is used $\partial_{j} () = \frac{\partial()}{\partial x^{j}}$ for the partial derivative and $d_{\nu} () = \frac{d()}{d x^{\nu}}$ for the total derivative with respect to the spatial components, namely: $x^{0}=t$, $x^{1}=x$, $x^{2}=y$, $x^{3}=z$.} in a generally curved space, given by
\begin{equation}\label{eq:15}
    (\nabla f)^{i} = \frac{g^{ij}}{\sqrt{|g|}} \partial_{j}f,
\end{equation}
where $g^{ij}$ are the spatial components of the inverse of the metric tensor $g_{\mu \nu}$ which, in general relativity, is a tensor of order $2$, not always positive definite that offers all the information about the causal structure and geometry of the space-time of interest, from which various quantities are also defined, namely: distance, volume, angle, past, future, curvature, among others. Furthermore, it is important to highlight that in Eq. (\ref{eq:15}) we have $|g| = |\det(g_{\mu \nu})|$. Then, we can rewrite expressions for the potential gradients as
\begin{equation}\label{eq:16}
    (\nabla V)^{i} = \frac{g^{ij}}{\sqrt{|g|}} \partial_{j}V
\end{equation}
and
\begin{equation}\label{eq:17}
    (\nabla Q)^{i} = \frac{g^{ij}}{\sqrt{|g|}} \partial_{j}Q.
\end{equation}
Furthermore, the mathematical expression for acceleration in a general system of orthogonal coordinates is given by
\begin{equation}
    \ddot{x}^{i} = d_{t}^{2}x^{i} + G^{i}_{jk}\partial x^{j} \partial x^{k} = d_{t}^{2}x^{i} + \omega^{i}.\label{eq:18}
\end{equation}
Above $G_{jk}^{i}$ are the Christoffel symbols in the orthogonal coordinate system. Then, considering the expressions (\ref{eq:16}), (\ref{eq:17}) and (\ref{eq:18}), and knowing that $\ddot{x}^{i}$ denotes the acceleration of the particle in the $i$-th spatial direction, calculated in curvilinear coordinates, equation (\ref{eq:12}) is reformulated as:
\begin{eqnarray}\label{eq:19}
     &\Gamma^{i}_{j k}d_{t}x^{j} d_{t}x^{k} - 2c\Gamma^{i}_{0j} d_{t}x^{j} + c^{2}\Gamma^{i}_{00} = \omega^{i} + \frac{g^{ij}}{m\sqrt{|g|}} \partial_{j}V + \frac{g^{ij}}{m\sqrt{|g|}} \partial_{j}Q.
\end{eqnarray}
Next, considering the general expression for the momentum components, $(\vec{P})^{i} = P^{i} =\hbar (\nabla S)^{i}$, which, similarly to Eqs. (\ref{eq:16}) and (\ref{eq:17}), can be rewritten as
\begin{equation}\label{eq:20}
    P^{i} = \frac{ \hbar g^{ij}}{\sqrt{|g|}} \partial_{j}S \;\; \rightarrow \;\; d_{t}x^{i} = \frac{\hbar g^{ij}}{m\sqrt{|g|}} \partial_{j}S.
\end{equation}
Using the Christoffel symbols
\begin{eqnarray}\label{eq:21}
    &\Gamma^{i}_{j k}&= \frac{g^{ih}}{2} \{ \partial_{j} g_{hk} + \partial_{k} g_{jh} - \partial_{h} g_{jk} \}, \nonumber \\
    &\Gamma^{i}_{j 0}&= \frac{g^{ih}}{2} \{ \partial_{j} g_{h0} + \partial_{0} g_{jh} - \partial_{h} g_{j0} \}, \\
    &\Gamma^{i}_{0 0}&= \frac{g^{ih}}{2} \{ 2 \partial_{0} g_{h0} - \partial_{h} g_{00} \}, \nonumber
\end{eqnarray}
the expression (\ref{eq:19}) is recast as follows:
\begin{eqnarray}\label{eq:22}
    &&\frac{\hbar^{2} g^{ih}}{2(m\sqrt{|g|})^{2}} \{ \partial_{j} g_{hk} + \partial_{k} g_{jh} - \partial_{h} g_{jk} \}g^{jl}g^{kn} \partial_{l}S \partial_{n}S \nonumber \\
    &&- \frac{c \hbar g^{ih}}{m\sqrt{|g|}} \{ \partial_{j} g_{h0} + \partial_{0} g_{jh} - \partial_{h} g_{j0} \}g^{jl} \partial_{l}S \\
    &&+ \frac{c^{2}g^{ih}}{2} \{ 2\partial_{0} g_{h0} - \partial_{h} g_{00} \} = \omega^{i} + \frac{g^{il}}{m\sqrt{|g|}} \partial_{l}V + \frac{g^{il}}{m\sqrt{|g|}} \partial_{l}Q. \nonumber
\end{eqnarray}
The expression (\ref{eq:22}) represents a geometrodynamic extension of the quantum force in Eq. (\ref{eq:11}) \cite{Gomez201900}. In simple terms, it provides an extended mathematical formalism that allows the description of physical systems based on their geometric evolution, that is, ``mass-energy tells space-time how to curve and space-time tells mass-energy how to move''. Thus, what is obtained is a formalism where the central mathematical object of Bohm's theory (the wave function) maintains its prominent role \cite{Einstein192800}. Moreover, in this geometrodynamic extension, not only does the wave function preserve its central role as the entity encoding the system's dynamics, but, in this formalism, the particle and the wave function, which are well-defined and clearly distinct entities, assume a dialectical role  \cite{Gomez201900}.

\section{Construction of the General Solution to the Problem}
\label{sec:4}
In this section, we present the deduction of the Bohmian solution to the one-dimensional problem of the scattering of non-relativistic and massive particles incident from the left over a barrier of constant potential \cite{Sakurai201100,Zettili200900,Pena200600}, based on the direct application of the formalism introduced in the previous sections \ref{sec:2} and \ref{sec:3}. Specifically, what is presented is the construction of an ansatz composed of the solution for each of the regions of interest, offered by the formalism of the sections \ref{sec:2} and \ref{sec:3}, and related between them by the conditions of continuity and under the protection of the local equivalence relationship between the Euclidean space-time, where Bohm mechanics is defined, and the space-time of the Lorentz variety offered by Eq. (\ref{eq:13}).

\begin{figure}[t]
    \centering
    \includegraphics[scale=0.750]{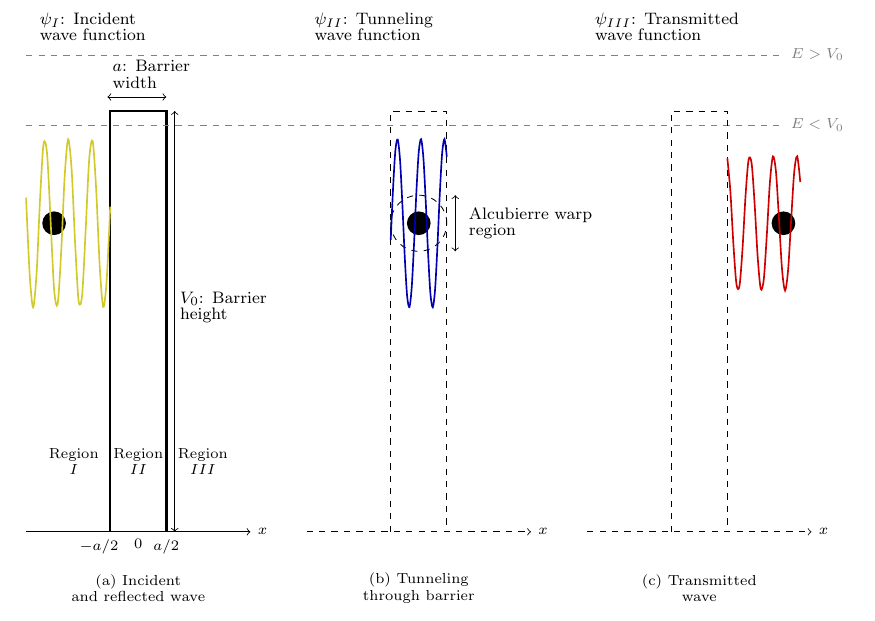}
    \caption{Schematic representation of the quantum tunneling phenomenon of a particle with energy $E$, incident from the left on a potential barrier of height $V_{0}$ and width $a$. Specifically, images (a), (b) and (c) are presented, where the regions of interest $I$, $II$ and $III$ are considered. In image (a), the incidence of a Bohmian particle in black and its guide wave in yellow on the potential barrier is presented in region $I$. In image (b), in region $II$ the Bohmian particle in black and its guiding wave in blue, inscribed within the hypothetical Alcubierre ball, a larger circumference with a discontinuous line, that is, the hypothetical representation of the tunnel effect as a distortion of space-time. Finally. In image (c) the transmitted Bohmian particle in black and its guide wave in red. In that sense, the regions $I$, $II$ and $III$ are considered in the construction of the general solution presented in section \ref{sec:4}.}
    \label{fig:Potential barrier}
\end{figure}

For this reason, we will consider the following three regions of interest. The first one is denoted as $I$, where the particle is incident, guided by its pilot wave, and subsequently reflected or not. The second region is denoted $II$, bounded by the potential barrier $V$, which has width $a$ and height $V_{0}$, which is mathematically defined as:
\begin{equation}\label{eq:23}        
    V(x) = V_{0} \Theta(x-a/2) \Theta(a/2-x).
\end{equation}
Here $\Theta$ is the Heaviside step function, which is mathematically defined as:
\[
    \Theta(x) =
    \left\{
    \begin{array}{cc}
        0 & \mbox{for $x < 0$} \\
        1 & \mbox{for $x \geq 0$}
    \end{array}.
    \right.
\]
In other words, it is the region where the quantum tunneling phenomenon occurs. Finally, the last region considered is denoted as $III$, where particle transmission, with its corresponding pilot wave, occurs. See Fig. \ref{fig:Potential barrier}.

So, the ansatz that describes the motion of particles in regions $I$ and $III$ must be of the form $\psi \propto \exp{\{ -i p_{\mu} x^{\mu} / \hbar \}} = \exp{\{ -i [kx - Et]/ \hbar \}} = \exp{\{ i S(x,t) \}}$, which satisfies the Klein-Gordon equation
\begin{equation}\label{eq:25}
    \biggl(\partial_{\mu} \partial^{\mu} - \frac{m_{0}^{2}c^{2}}{\hbar^{2}} \biggr) \psi = 0.
\end{equation}
The dispersion relation \cite{Greiner200000} is $E = \pm c \sqrt{m_{0}^{2}c^{2}+k^{2}}$.

In the non-relativistic limit, where $E^{\prime} = E - m_{0}c^{2}$, the wave function takes the form $\psi \propto \phi(x,t)\exp{\{-i m_{0}c^{2}t/ \hbar\}}$. This leads to the Schr\"odinger equation
\begin{equation}\label{eq:26}
    i \hbar \partial_{t} \phi = - \frac{\hbar^{2}}{2m_{0}} \partial_{x}^{2} \phi.
\end{equation}
Thus, the solutions in regions $I$ and $III$ are given by the expressions:
\begin{eqnarray}\label{eq:27}
    \psi_{I}(x,t) &=& A \exp{\{ ip_{\mu}x^{\mu}/\hbar \}} + B \exp{\{ -ip_{\mu}x^{\mu}/\hbar \}} \\
    &=& \sqrt{\rho_{I}} \exp{\{ i S_{I}(x,t) \}}\ \nonumber
\end{eqnarray}
for  $x \leq -a/2$ and
\begin{eqnarray}\label{eq:28}
    \psi_{III}(x,t) &=& F \exp{\{ ip_{\mu}x^{\mu}/\hbar \}} \\
    &=& \sqrt{\rho_{III}} \exp{\{ i S_{III}(x,t) \}} \, \nonumber
\end{eqnarray}
for $x \geq a/2$, where 
$ \sqrt{\rho_{I}} = \sqrt{\mathcal{A}^{2} \cos^{2}{\{ p_{\mu}x^{\mu}/\hbar \}} + \mathcal{B}^{2} \sin^{2}{\{ p_{\mu}x^{\mu}/\hbar \}}} $ with $\mathcal{A} = A + B$ and $\mathcal{B} = i(A - B)$, $ \sqrt{\rho_{III}} = F = \mathcal{F} $, $S_{I}(x,t) = \tan^{-1}{ \{ \mathcal{B} \tan{\{ p_{\mu}x^{\mu}/\hbar \}}/\mathcal{A} \} }$, and $S_{III}(x,t) = p_{\mu}x^{\mu}/\hbar$. 
To reconcile the experimental findings of Sharoglazova et al. \cite{Sharoglazova202500}—where finite dwell times were observed for particles tunneling into a barrier, contradicting the Bohmian prediction of infinite dwell times—the general solution in regions $I$ and $III$ must be expressed as a superposition of distinct energy eigenstates, as argued in \cite{Dewdney198700,Barut198800}:
\begin{equation}
    \Psi_{j}(x,t) = \sum_{l=1}^{2} C_{j}^{l} \psi_{j}^{l}(x,t) = \sum_{l=1}^{2} C_{j}^{l} \sqrt{\rho_{j}^{l}} \exp{ \{ i S_{j}^{l}(x,t) \} }, \label{eq:29}
\end{equation}
where $j=I$ for $x \leq -a/2$ and $j=III$ for $x \geq a/2$. Here, $S_{j}^{1}(x,t)$ and $S_{j}^{2}(x,t)$ correspond to phase functions associated with two distinct energy eigenvalues $E^{1}$ and $E^{2}$, respectively, introduced to model the interference and finite dwell time observed experimentally \cite{Sharoglazova202500}. This superposition is necessary to describe:
\begin{itemize}
    \item The interference between incident and reflected waves (Region I),
    \item Nontrivial transmission (Region III),
    \item And crucially, to ensure consistency with both probability conservation \textit{and} the observed energy-speed relationship $v = \sqrt{2|\Delta|/m}$ from \cite{Sharoglazova202500}, where faster motion corresponds to more negative local kinetic energy.
\end{itemize}
Rewriting the solution in the Bohmian form reveals how finite velocities emerge:
\begin{equation}\label{eq:30}
    S_{j}(x,t) = \tan^{-1}{ \left( \frac{ \sum_{l=1}^{2} C_{j}^{l} \sqrt{\rho_{j}^{l}} \sin{ S_{j}^{l}(x,t) } }{ \sum_{l=1}^{2} C_{j}^{l} \sqrt{\rho_{j}^{l}} \cos{ S_{j}^{l}(x,t) } } \right)},
\end{equation}
with the density:
\begin{equation}\label{eq:31}
    \sqrt{\rho_{j}} = \sqrt{ \sum_{l=1}^{2} (C_{j}^{l})^{2}\rho_{j}^{l} + 2 C_{j}^{q}C_{j}^{q+1}\sqrt{ \rho_{j}^{q}\rho_{j}^{q+1} } \cos{ \Xi_{j}^{q} } },
\end{equation}
with $\Xi_{j}^{q} = S_{j}^{q+1}(x,t) - S_{j}^{q}(x,t)$ and $q=1$.

The phase gradients in (\ref{eq:30}) break the Bohmian stasis paradox—where single eigenstates predict stationary particles in the barrier—by introducing non-zero velocity fields via $\nabla S_{j}(x,t)$. This resolves the discrepancy with experimental results while preserving the causal structure of Bohmian mechanics.

Therefore, knowing the wave functions for our problem of interest in regions $I$ and $III$, in the context of geometrodynamics it is necessary to know the form of the metric that describes spacetime. Considering $\omega^{i} = 0$ for $i = 1,2,3$ (a spacetime with flat spatial geometry) and $\partial_{l} V = 0$ (for the case of the constant potential barrier considered in this work) in Eq. (\ref{eq:22}), the following expressions are obtained
\begin{eqnarray}
    0 &=& \frac{c^{2}g^{10} \partial_{0}g_{00}}{2} + \frac{c^{2}g^{11} \{ 2\partial_{0}g_{10} - \partial_{1}g_{00} \}}{2}  \frac{g^{11} \partial_{1}Q}{m\sqrt{|g|}} \;\; \{\mbox{for} \; i=1,\ j \neq 1\}, \nonumber \\
    0 &=& \frac{c \hbar g^{10} (\partial_{1} g_{00}) \partial_{1}S}{m\sqrt{|g|}} \;\; \{\mbox{for} \; i=j=1,\ k \neq 1\}, \; \label{eq:32} \\
    0 &=& \frac{\hbar^{2} g^{01} (2 \partial_{1} g_{01}) (g^{11})^{2}(\partial_{1}S)^{2}}{2(m\sqrt{|g|})^{2}}  \frac{c \hbar g^{10} (\partial_{1} g_{00}) g^{11} \partial_{1}S}{m\sqrt{|g|}} \;\; \{\mbox{for} \; i=j=k = 1\}. \nonumber
\end{eqnarray}
Then, from Eq. (\ref{eq:32}), it is deduced that the most general metric that describes the spacetime in these regions is of the form
\begin{equation}\label{eq:33}
    g_{\mu \nu} =
    \left(
    \begin{array}{cc}
        - \left\{\frac{3Q}{mc^{2}}\right\}^{\frac{2}{3}} & \mathcal{O}_{1\times3} \\
        \mathcal{O}_{3\times1} & \mathcal{I}_{3\times3}
    \end{array}
    \right),
\end{equation}
whose inverse tensor is given by the expression
\begin{equation}\label{eq:34}
    g^{\mu \nu} =
    \left(
    \begin{array}{cc}
        - \left\{\frac{3Q}{mc^{2}}\right\}^{-\frac{2}{3}} & \mathcal{O}_{1\times3} \\
        \mathcal{O}_{3\times1} & \mathcal{I}_{3\times3}
    \end{array}
    \right),
\end{equation}
where $\mathcal{I}_{3\times3}$ is the identity matrix of order $3\times3$, $\mathcal{O}_{3\times1}$ the null column matrix of order $3\times1$ and $\mathcal{O}_{ 1\times3}$ is the null line matrix elements of order $1\times3$.

On the other hand, in region $II$ the metric that describes spacetime must be of the Alcubierre type \cite{Alcubierre199400}, of the Krasnikov type  \cite{Krasnikov199800}, or some type of metric that describes a hyperbolically separable spacetime, since these types of metrics support particle travel at hyperluminous speeds \cite{Alcubierre200500}. In this sense, in the present work we consider the Alcubierre-like metric, given by
\begin{equation}\label{eq:35}
    \mathcal{G}_{\mu \nu} =
    \left(
    \begin{array}{cc}
        \mbox{\ss}_{k} \mbox{\ss}^{k} - 1 & - \mbox{\ss}_{i} \\
        - \mbox{\ss}_{j} & \gamma_{ij}
    \end{array}
    \right),
\end{equation}
which has the form of the metric reported in the literature  \cite{Foo202200}, 
and whose inverse tensor is given by the expression
\begin{equation}\label{eq:36}
    \mathcal{G}^{\mu \nu} =
    \left(
    \begin{array}{cc}
        -1 & - \mbox{\ss}^{i} \\
        - \mbox{\ss}^{j} & \gamma^{ij} - \mbox{\ss}^{i} \mbox{\ss}^{j}
    \end{array}
    \right),
\end{equation}
where $i,j,k = 1,2,3$ are spatial indices, $\mbox{\ss}_{i} = v_{s}(t) f(r_{s})$ for $i = 1$ and $\mbox{\ss}_{i} = 0$ otherwise defines the spacetime coupling along $x$, $\gamma_{ij} = \mathrm{diag}(1,1,1)$ is the flat spatial metric, $r_{s}(t) \equiv x - x_{s}(t)$ denotes the position relative to the bubble's center, $v_{s}(t) = d_{t}[x_{s}(t)] = \dot{x}_{s}(t)$ represents the bubble's instantaneous velocity, $f(r_{s}) = \alpha_{0}/\{ v_{s} [1 + \alpha_{1} \cosh{(2 \sigma r_{s})}] \}$ characterizes the bubble profile. And where:
\begin{itemize}
    \item $\alpha_{0} = - [v_{s} \tanh{(2 \sigma R)}]/[2 \tanh{(\sigma R)}]$ governs the peak distortion strength;
    \item $\alpha_{1} = \mathrm{sech}{(2 \sigma R)}$ controls the spatial decay rate;
    \item $\sigma$ determines the distortion region's width;
    \item $R$ is the effective radius of the Alcubierre bubble, defining the spatial extent where the metric distortion is non-negligible ($|r_{s}| \lesssim R$).
\end{itemize}
Furthermore, the function $f(r_{s})$ peaks at $\alpha_{0}/[v_{s}(1+\alpha_{1})]$ when $r_{s}=0$ (bubble center), with $\alpha_{0}$ setting the maximum distortion amplitude. For $|r_{s}| > 0$, $f(r_{s})$ decays exponentially, becoming negligible at $|r_{s}| \gg R$.

Therefore, $g_{\mu \nu}$ and $\mathcal{G}_{\mu \nu}$ allows us to write the metric piece-wise
\[
    \hat{g}_{\mu \nu} =
    \left\{
    \begin{array}{ccc}
        g_{\mu \nu} &\mbox{for}& x \leq -a/2, \\
        \mathcal{G}_{\mu \nu} &\mbox{for}& -a/2 < x < a/2, \\
        g_{\mu \nu} &\mbox{for}& x \geq a/2,
    \end{array}
    \right.
\]
which is clearly continuous at the boundaries, since, at the entry points $\mathcal{P}_{0}(-a/2,t_{0})$ and exit points $\mathcal{P}_{1}(a/2,t_{1})$ the perturbation velocity must be null ($v_{s}$), therefore $\mathcal{G}_{\mu \nu} = \mbox{diag}=(-1,1,1,1)$. Furthermore, by construction we know that $g_{\mu \nu}$ reduces to a Minkowski-type metric when $Q \rightarrow mc^{2}/3 = 1/3$ (recalling that in this article $c = 1$, $\hbar = 1$ and $m = 1$) for the description to be consistent. Thus, $g_{\mu \nu} = \mathcal{G}_{\mu \nu} = \mbox{diag}=(-1,1,1,1)$ at the contact points between the regions, which ultimately describes the entire spacetime through which the particles propagate.

Next, by considering the nontrivial solutions of Eq. (\ref{eq:32}), the expression
\begin{equation}\label{eq:38}
    dS = - 2 dr_{s} v_{s}f \left\{ \frac{1}{1 - v^{2}_{s}f^{2}} \right\}
\end{equation}
is deduced, which allows calculating the phase function in region $II$. Similarly, the expression
\begin{equation}\label{eq:39}
    dQ = df \left\{ \frac{v_{s}}{1 - v_{s}^{2}f^{2}} + v_{s}^{2}f \right\}
\end{equation}
is obtained, whose solution will be presented later (see Section \ref{sec:5}), allowing us to obtain the expression for the quantum potential in region $II$. Thus, by integrating Eq. (\ref{eq:38}), as shown in the part $I$ of the Appendix, the phase function is obtained:
\begin{eqnarray}\label{eq:40}
    &S_{II}(x,x_{s}(t)) = \frac{ \alpha_{0} \beta_{0}}{2 \sigma \sqrt{(\beta_{3})^{2} - (\beta_{2})^{2}}} \Biggl\{ \frac{\beta_{1} - \mu_{0}}{\sqrt{\mu_{0}}} \tan^{-1}{\Biggl[\frac{\tan{(x^{\prime}/2)}}{\sqrt{\mu_{0}}} \Biggr]} \\
    &- \frac{\beta_{1} - \mu_{1}}{\sqrt{\mu_{1}}} \tan^{-1}{\Biggl[\frac{\tan{(x^{\prime}/2)}}{\sqrt{\mu_{1}}} \Biggr]} \Biggr\}. \nonumber
\end{eqnarray}
Above $\beta_{0} = [2(\alpha_{1} - 1)]/[(\alpha_{1} - 1)^{2} - \alpha^{2}_{0}]$, $\beta_{1} = (\alpha_{1} + 1)/(\alpha_{1} - 1)$, $\beta_{2} = [(\alpha_{1} + 1)^{2} - \alpha^{2}_{0}]/[(\alpha_{1} - 1)^{2} - \alpha^{2}_{0}]$, $\beta_{3} = [\alpha_{1}^{2} - 1 + \alpha^{2}_{0}]/[(\alpha_{1} - 1)^{2} - \alpha^{2}_{0}]$, $\mu_{0} = \beta_{3} - \sqrt{(\beta_{3})^{2} - (\beta_{2})^{2}}$, $\mu_{1} = \beta_{3} + \sqrt{(\beta_{3})^{2} - (\beta_{2})^{2}}$ and
\begin{equation}\label{eq:41}
    x' = \cos^{-1}{ \left[ \mbox{sech}{(2 \sigma r_{s})} \right] }.
\end{equation}
Finally, we can write the phase function as follows
\begin{equation}\label{eq:42}
    S_{II}^{l} (x,t) = S_{II}(x,x_{s}(t)) - (V_{0} - E^{l})t,
\end{equation}
which allows us to find the general solution in region $II$, given by the superposition
\begin{equation}\label{eq:43}
    \Psi_{II}(x,t) = \sum_{l=1}^{2} C_{II}^{l} \exp{ \{ i S_{II}^{l}(x,t) \} },
\end{equation}
where $E^{1}$ and $E^{2}$ in (\ref{eq:43}) are two distinct energy values used in the superposition to model the tunneling dynamics. However, it is important to note that physically acceptable solutions for Eq. (\ref{eq:43}), leading to a consistent phase function (\ref{eq:40}), are those where $\alpha_{1} \neq 0$ or $\alpha_{1} \neq 1$. Now, to be able to construct a general solution to the problem, it is necessary to impose continuity conditions that are consistent with the physics of the problem.

For this reason, it is essential to emphasize that the phases $S_{j}$ and the amplitudes $\sqrt{\rho_{j}}$ defined by expressions (\ref{eq:29})-(\ref{eq:30}) are functions that depend only on $x^{\mu}$, and therefore transform as scalars $S_{j}^{\prime}(x^{\mu \prime}) = S_{j}(x^{\mu})$ and $\rho_{j}^{\prime}(x^{\mu \prime}) = \rho_{j}(x^{\mu})$. So, under a coordinate transformation $x \rightarrow x^{\mu}$, $\Psi_{j}$ maintains its form, since $S_{j}$ and $\sqrt{\rho_{j}}$ are invariant. Similarly, the phase $S_{II}$ defined by (\ref{eq:40}) is derived from the metric $\mathcal{G}{\mu \nu}$ of expression (\ref{eq:34}), resulting in a scalar function that evidently transforms as a scalar; moreover, it depends on $r_{s}$, which is invariant.

On the other hand, the covariant derivative\footnote{Following the line of clarifications made previously in sections \ref{sec:2} and \ref{sec:3}, we observe that in this work $()_{;\nu} = \nabla_{\nu} () =\partial_{\nu} () + \Gamma_{\mu \nu}^{\rho}()$ is used to denote the covariant derivative.} $\Psi_{j;\mu} = \partial_{\mu} \Psi_{j} + \Gamma_{\mu \nu}^{\rho}\Psi_{j}$, must be continuous at the boundaries. This is easily verifiable by recalling that for $j =I,III$ the symbols $\Gamma_{\mu \nu}^{\rho}$ are non-null only for $\mu=\rho=1$, $\nu = 0$, which leads to $\Gamma_{1 0}^{1} = \partial_{0}g_{00}/2 = 0$ because $g_{00} = - \{3Q\}^{2/3}$ is constant, making the covariant derivative reduce to $\Psi_{j;x} = \partial_{x} \Psi_{j}$. In region $II$, the non-null $\Gamma_{\mu \nu}^{\rho}$ symbols (when $-a/2 < x < a/2$) are $\Gamma_{0 0}^{1} = v_{s}^{2} f \partial_{1}f/2$, $\Gamma_{0 1}^{0} = \Gamma_{1 0}^{1} = - v_{s} \partial_{1}f/2$, and as already stated, they are all null at the boundary since $v_{s} = 0$, so the covariant derivative again reduces to $\Psi_{j;x} = \partial_{x} \Psi_{j}$.

Then, imposing the continuity conditions of the solution and its derivatives, the expressions obtained from Eqs. (\ref{eq:29}) and (\ref{eq:43}) must be substituted in
\begin{eqnarray}
    &\Psi_{k} (x,t) \biggl|_{(x_{k}, t_{k})} = \Psi_{k+I} (x,t) \biggl|_{(x_{k}, t_{k})}, \\
    &[\partial_{x} \Psi_{k} (x,t)] \biggl|_{(x_{k}, t_{k})} = [\partial_{x} \Psi_{k+I} (x,t)] \biggl|_{(x_{k}, t_{k})}, \label{eq:44}
\end{eqnarray}
with $x_{k} \in [-a/2, a/2]$, $t_{k} \in [t_{0}, t_{1}]$ (where $t_0$ and $t_1$ mark the particle's arrival at the boundaries $x = -a/2$ (entry) and $x = a/2$ (exit), respectively) and $k = I, II, III$, we obtain the system of equations:
\begin{eqnarray}
    \mathcal{C}_{I}^{1} \zeta_{1} + \mathcal{C}_{I}^{2} \zeta_{2} e^{i\Delta E t_{0}} &=& \zeta_{3} [ \mathcal{C}_{II}^{1} + \mathcal{C}_{II}^{2} e^{i\Delta E t_{0}} ], \nonumber \\
    \mathcal{C}_{I}^{1} \zeta_{1}^{\prime} + \mathcal{C}_{I}^{2} \zeta_{2}^{\prime} e^{i\Delta E t_{0}} &=& \zeta_{3}^{\prime} [ \mathcal{C}_{II}^{1} + \mathcal{C}_{II}^{2} e^{i\Delta E t_{0}} ], \label{eq:45} \\
    \mathcal{C}_{III}^{1} \zeta_{5} + \mathcal{C}_{III}^{2} \zeta_{6} e^{i\Delta E t_{1}} &=& \zeta_{4} [ \mathcal{C}_{II}^{1} + \mathcal{C}_{II}^{2} e^{i\Delta E t_{1}} ], \nonumber \\
    \mathcal{C}_{III}^{1} \zeta_{5}^{\prime} + \mathcal{C}_{III}^{2} \zeta_{6}^{\prime} e^{i\Delta E t_{1}} &=& \zeta_{4}^{\prime} [ \mathcal{C}_{II}^{1} + \mathcal{C}_{II}^{2} e^{i\Delta E t_{1}} ]. \nonumber
\end{eqnarray}
Consequently, by solving the system of equations (\ref{eq:45}) while imposing the superposition condition $\mathcal{C}_{II}^{2} = 1 - \mathcal{C}_{II}^{1}$, the expressions for the coefficients are obtained:
\begin{eqnarray}\label{eq:46}
    \mathcal{C}_{I}^{1} &=& \varpi_{4} e^{-i\Delta E t_{0}} [(e^{i\Delta E t_{0}} - 1) \mathcal{C}_{II}^{1} + 1], \nonumber \\
    \mathcal{C}_{I}^{2} &=& \varpi_{3} [(e^{i\Delta E t_{0}} - 1) \mathcal{C}_{II}^{1} + 1], \nonumber \\
    \mathcal{C}_{II}^{1} &=& - \frac{e^{i\Delta E t_{1}}}{e^{i\Delta E t_{1}} - 1} \left\{ \varpi_{2} \mathcal{C}_{III}^{1} - 1 \right\}, \\
    \mathcal{C}_{II}^{2} &=& \frac{1}{e^{i\Delta E t_{1}} - 1} \left\{ \varpi_{2} e^{i\Delta E t_{1}} \mathcal{C}_{III}^{1} - 1 \right\}, \nonumber \\
    \mathcal{C}_{III}^{2} &=& \varpi_{1} e^{i\Delta E t_{1}} \mathcal{C}_{III}^{1}. \nonumber
\end{eqnarray}
Above we defined $\varpi_{1} = [\zeta_{4}^{\prime}\zeta_{5} - \zeta_{4}\zeta_{5}^{\prime}]/[\zeta_{4}\zeta_{6}^{\prime} - \zeta_{4}^{\prime}\zeta_{6}]$, $\varpi_{2} = [\zeta_{5}^{\prime}\zeta_{6} - \zeta_{5}\zeta_{6}^{\prime}]/[\zeta_{4}\zeta_{6}^{\prime} - \zeta_{4}^{\prime}\zeta_{6}]$, $\varpi_{3} = [\zeta_{1}\zeta_{3}^{\prime} - \zeta_{1}^{\prime}\zeta_{3}]/[\zeta_{1}\zeta_{2}^{\prime} - \zeta_{1}^{\prime}\zeta_{2}]$ and $\varpi_{4} = [\zeta_{2}^{\prime}\zeta_{3} - \zeta_{2}\zeta_{3}^{\prime}]/[\zeta_{1}\zeta_{2}^{\prime} - \zeta_{1}^{\prime}\zeta_{2}]$ with
\begin{eqnarray}
    \zeta_{l} &=& \sqrt{\rho_{I}^{l}(x,t)}\exp{\{iS_{I}^{l}(x,t)\}} \biggl|_{(- \frac{a}{2},t_{0})}, \nonumber \\
    \zeta_{l}^{\prime} &=& \frac{k^{l}[(\mathcal{B}^{2} - \mathcal{A}^{2}) \sin{\{ 2p_{\mu}x^{\mu} \}} + i2 \mathcal{A} \mathcal{B}]}{2 \sqrt{\rho_{I}^{l}(x,t)}}  \exp{\{ iS_{I}^{l}(x,t) \}} \biggl|_{(- \frac{a}{2},t_{0})}, \nonumber \\
    \zeta_{l+2} &=& \exp{\{i [S_{II}(x,x_{s}(t)) - V_{0}t]\}} \biggl|_{((-1)^{l+2}\frac{a}{2},t_{0})}, \nonumber \\
    \zeta_{l+2}^{\prime} &=& \frac{i \alpha_{0} \beta_{0} \sec^{2}{(\frac{x^{\prime}}{2})} \tan{(2 \sigma r_{s})}}{2 \sqrt{(\beta_{3})^{2} - (\beta_{2})^{2}} \sqrt{\cosh^{2}{(2 \sigma r_{2})} - 1}} \\
    &\times & \exp{\{i[S_{II}(x,x_{s}(t)) - V_{0}t]\}} \nonumber \\
    &\times& \biggl[ \frac{\beta_{1} - \mu_{1}}{\mu_{1} + \tan^{2}{(\frac{x^{\prime}}{2})}} - \frac{\beta_{1} - \mu_{0}}{\mu_{0} + \tan^{2}{(\frac{x^{\prime}}{2})}} \biggr] \biggl|_{((-1)^{l+2}\frac{a}{2},t_{0})}, \nonumber \\
    \zeta_{l+4} &=& \exp{\{iS_{III}^{l}(x,t)\}} \biggl|_{(\frac{a}{2},t_{0})}, \nonumber \\
    \zeta_{l+4}^{\prime} &=& ik^{l} \exp{\{iS_{III}^{l}(x,t)\}} \biggl|_{(\frac{a}{2},t_{0})}. \nonumber
\end{eqnarray}
In the expressions above, values for $l=1,2$ are considered, along with the energy variation $\Delta E = E^{2} - E^{1}$. Additionally, the coefficients were rewritten as $C_{j}^{l} = \mathcal{C}_{j}^{l}$ if $j = I,II$ and $C_{j}^{l} \sqrt{\rho_{j}^{l}} = \mathcal{C}_{j}^{l}$ for $j=III$, again for all $l=1,2$.

In summary, through the coefficients presented in Eq. (\ref{eq:46}) and the expressions for $S_{I}^{l}$, $S_{II}$, and $S_{III}^{l}$, the general solution to the problem for each of the regions of interest can be written as:
\begin{eqnarray}\label{eq:48}
    \Psi_{I}(x,t) &=& \mathcal{C}_{I}^{1} \sqrt{\rho_{I}^{1}} \exp{ \{ iS_{I}^{1}(x,t) \} } + \mathcal{C}_{I}^{2} \sqrt{\rho_{I}^{2}} \exp{ \{ iS_{I}^{2}(x,t) \} }, \\
    \Psi_{II}(x,t) &=& [ \mathcal{C}_{II}^{1} \exp{ \{ i E^{1} t \} } + \mathcal{C}_{II}^{2} \exp{ \{ i E^{2} t \} } ]  \exp{ \{ i [S_{II}(x,x_{s}(t)) - V_{0}t] \} }, \nonumber  \\
    \Psi_{III}(x,t) &=& \mathcal{C}_{III}^{1} \exp{ \{ iS_{III}^{1}(x,t) \} } + \mathcal{C}_{III}^{2} \exp{ \{ iS_{III}^{2}(x,t) \} }, \nonumber
\end{eqnarray}
where, as expected, the general expression is written in terms of the geometry of the spacetime considered by imposing the geometrodynamic constraint (\ref{eq:13}). Then, by substituting $\mathcal{C}_{III}^{2}$ into the expression for $\sqrt{\rho_{III}}$, the transmission coefficient is obtained:
\begin{equation}\label{eq:49}
    T = \mathcal{C}_{III}^{1} \sqrt{ 1 + \varpi_{1}^{2} e^{i 2 \Delta E t_{1}} + 2 \varpi_{1} e^{i \Delta E t_{1}} \cos{\Xi_{III}} }.
\end{equation}
\section{General Expression for the Quantum Potential}
\label{sec:5}
In this section, we apply the formalism and results presented in Sections \ref{sec:3} and \ref{sec:4} to derive the general expression of the quantum potential for each of the regions considered in the problem of interest.

For this purpose, it is necessary to recall that the metric describing spacetime in regions $I$ and $III$ implies that the quantum potential in these regions must be of the form $Q_{j} = \pm mc^{2} \{ g_{00} \}^{3/2}/3 = \pm \{ g_{00} \}^{3/2}/3$, where again $j = I$ as $x \leq -a/2$ and $j = III$ as $x \geq a/2$. However, as mentioned in the previous section, the quantum potential in region II is determined by integrating Eq. (\ref{eq:39}), resulting in the expression
\begin{equation}\label{eq:50}
    Q_{II}(x,t) = \Biggl\{ \ln{\Biggl| \frac{1 + v_{s}f}{\sqrt{1 - v^{2}_{s}f^{2}}} \Biggr|} + \frac{v^{2}_{s}f^{2}}{2} \Biggr\}.
\end{equation}
In essence, particles within the barrier, where quantum tunneling occurs and the Hartman effect is exhibited, are influenced by a quantum potential that distorts spacetime.

\section{Momentum, Position, and Tunneling Time}
\label{sec:6}
In the previous sections, along with the theoretical foundations and the general solution, we derived the mathematical expressions for the quantum potential in each of the regions of interest in our problem. Even though the wave function and the quantum potential are fundamental quantities within the framework of Bohmian mechanics \cite{Barut198800,Dewdney198700,Oriols201900,Pena200600}, it is essential to determine the particle's momentum and position. Therefore, this section is dedicated to calculating the mathematical expressions for these quantities, which ultimately allows us to determine the quantum tunneling time \cite{Hartman196200,Winful200200, Winful200300, Buttiker198300,  Winful200301, Winful200400, Winful200600, Hagmann199300}.

In this regard, obtaining an expression for the linear momentum for the region $I$ turns out to be a difficult task due to the form of the general solution. Nevertheless, by rewriting $\sqrt{\rho_{I}^{l}}$ in the form
\begin{eqnarray}
    \sqrt{\rho_{I}^{l}} &=& \sqrt{\mathcal{A}^{2} \cos^{2}{\{ p_{\mu}x^{\mu}/\hbar \}} + \mathcal{B}^{2} \sin^{2}{\{ p_{\mu}x^{\mu}/\hbar \}}} \label{eq:51} \\ 
    &=& \sqrt{2AB + [A^{2} + B^{2}]\cos{\{ 2p_{\mu}x^{\mu} \}}} \nonumber
\end{eqnarray}
and considering the solution for low energies and reasonably small times ($p_{\mu} x^{\mu} \approx 0 \; \rightarrow \; \cos{\{
2 p_{\mu}x^{\mu} \}} \approx 1$), the expression (\ref{eq:51}) is simplified to a more direct form given by
\begin{eqnarray}\label{eq:52}
    \sqrt{\rho_{I}^{l}} &=& \sqrt{2AB + [A^{2} + B^{2}]\cos{\{ 2p_{\mu}x^{\mu} \}}} \\
    &=& \sqrt{(A + B)^{2}} \nonumber \\
    &=& \mathcal{A}. \nonumber
\end{eqnarray}
On the other hand, in the low-energy limit ($E \to 0$), the phase function $S_{I}^{l}$ simplifies to $S_{I}^{l}(x,t) = \tan^{-1}\left( \{\mathcal{B}\tan(p_{\mu}x^{\mu})/\mathcal{A}\} \right) \approx \mathcal{B}(k^{l}x - E^{l}t)/\mathcal{A}$,
where $p_{\mu}x^{\mu} = k^{l}x - E^{l}t$ and the approximation holds for small arguments. When substituted into the superposition (\ref{eq:30}), this yields the emergent phase function $S_{I}(x,t)$, whose gradient governs the Bohmian velocity field in the low-energy regime.

To determine the linear momentum in region $I$, we choose the metric in this region to be of the Minkowski type, i.e., $Q_{I} = mc^{2}/3$, which implies that Eq. (\ref{eq:20}) reduces to Eq. (\ref{eq:7}) when the particles incident on the potential barrier have non-relativistic velocities ($v \ll 1$, in natural units $\hbar = c = 1$), which is precisely our case of study. Then, substituting $S_{I}(x,t)$ into Eq. (\ref{eq:7}), we obtain
\begin{eqnarray}\label{eq:53}
    P_{I}(x,t) &=& \frac{\mathcal{B} [(\mathcal{C}_{I}^{1})^{2} k^{1} + (\mathcal{C}_{I}^{2})^{2} k^{2} + \mathcal{C}_{I}^{1} \mathcal{C}_{I}^{2} (k^{1} + k^{2}) \cos{\Xi_{I}}] }{\mathcal{A}[(\mathcal{C}_{I}^{1})^{2} + (\mathcal{C}_{I}^{2})^{2} + 2 \mathcal{C}_{I}^{1} \mathcal{C}_{I}^{2} \cos{\Xi_{I}}]}.
\end{eqnarray}
Analogously, rewriting equation (\ref{eq:29}) in terms of $S_{III}^{l}$ and $C_{III}^{l} \sqrt{\rho_{III}^{l}} = \mathcal{C}_{III}^{l}$, and substituting into Eq. (\ref{eq:7}), we determine that the linear momentum in region $III$ is given by
\begin{eqnarray}\label{eq:54}
    P_{III}(x,t) &=& \frac{(\mathcal{C}_{III}^{1})^{2} k^{1} + (\mathcal{C}_{III}^{2})^{2} k^{2} + \mathcal{C}_{III}^{1} \mathcal{C}_{III}^{2} (k^{1} + k^{2}) \cos{\Xi_{III}}}{(\mathcal{C}_{III}^{1})^{2} + (\mathcal{C}_{III}^{2})^{2} + 2 \mathcal{C}_{III}^{1} \mathcal{C}_{III}^{2} \cos{\Xi_{III}}}.
\end{eqnarray}
Alternatively, by substituting equation (\ref{eq:36}) into equation (\ref{eq:20}), where $g^{11} = [1 - \alpha_{0}^{2} + 2 \alpha_{1} \cosh{(2 \sigma r_{s})} + \alpha_{1}^{2} \cosh^{2}{(2 \sigma r_{s})}]/[1 + 2 \alpha_{1} \cosh{(2 \sigma r_{s})} + \alpha_{1}^{2} \cosh^{2}{(2 \sigma r_{s})}]$ and $\sqrt{|g|} = 1$, it is obtained that the linear momentum in region $II$ is given by
\begin{eqnarray}\label{eq:55}
    P_{II}(x,t) &=& \frac{ \alpha_{0} \beta_{0} \mbox{sech}(2 \sigma r_{s}) }{2 \sqrt{(\beta_{3})^{2} - (\beta_{2})^{2}}} \\
    &\times& \Big\{\frac{1 - \alpha_{0}^{2} + 2 \alpha_{1} \cosh{(2 \sigma r_{s})}}{1 + 2 \alpha_{1} \cosh{(2 \sigma r_{s})} + \alpha_{1}^{2} \cosh^{2}{(2 \sigma r_{s})}} \nonumber \\
    &+& \frac{\alpha_{1}^{2} \cosh^{2}{(2 \sigma r_{s})}}{1 + 2 \alpha_{1} \cosh{(2 \sigma r_{s})} + \alpha_{1}^{2} \cosh^{2}{(2 \sigma r_{s})}}\Big\} \nonumber \\
    &\times&\Biggl\{ \frac{ \beta_{1} - \mu_{0} }{\mu_{0} \cos^{2}{[\frac{\mbox{sech}(2 \sigma r_{s})}{2}]} + \sin^{2}{[\frac{\mbox{sech}(2 \sigma r_{s})}{2}]}} \nonumber \\
    &-& \frac{\beta_{1} - \mu_{1}}{\mu_{1}\cos^{2}{[\frac{\mbox{sech}(2 \sigma r_{s})}{2}]} + \sin^{2}{[\frac{\mbox{sech}(2 \sigma r_{s})}{2}]}} \Biggr\}. \nonumber
\end{eqnarray}
As a result, the following differential equations are derived from the expressions (\ref{eq:53}) and (\ref{eq:54}):
\begin{equation}\label{eq:56}
    \frac{dx}{dt} = \frac{\mathcal{B}}{\mathcal{A}} \frac{\vartheta_{3} \{ 1 + \vartheta_{4} \cos{\Xi_{I}} \} }{\vartheta_{1} \{ 1 + \vartheta_{2} \cos{\Xi_{I}} \} },
\end{equation}
\begin{equation}\label{eq:57}
    \frac{dx}{dt} = \frac{\vartheta_{7} \{ 1 + \vartheta_{8} \cos{\Xi_{III}} \} }{\vartheta_{5} \{ 1 + \vartheta_{6} \cos{\Xi_{III}} \} },
\end{equation}
where $\vartheta_{1}=(\mathcal{C}_{I}^{1})^{2} + (\mathcal{C}_{I}^{2})^{2}$, $\vartheta_{2}=2 \mathcal{C}_{I}^{1} \mathcal{C}_{I}^{2}/[(\mathcal{C}_{I}^{1})^{2} + (\mathcal{C}_{I}^{2})^{2}]$, $\vartheta_{3}=(\mathcal{C}_{I}^{1})^{2}k^{1} + (\mathcal{C}_{I}^{2})^{2}k^{2}$, $\vartheta_{4}=\mathcal{C}_{I}^{1} \mathcal{C}_{I}^{2}(k^{1}+k^{2})/[(\mathcal{C}_{I}^{1})^{2}k^{1} + (\mathcal{C}_{I}^{2})^{2}k^{2}]$, $\vartheta_{5}=(\mathcal{C}_{III}^{1})^{2} + (\mathcal{C}_{III}^{2})^{2}$, $\vartheta_{6}=2\mathcal{C}_{III}^{1} \mathcal{C}_{III}^{2}/[(\mathcal{C}_{III}^{1})^{2} + (\mathcal{C}_{III}^{2})^{2}]$, $\vartheta_{7} = (\mathcal{C}_{III}^{1})^{2}k^{1} + (\mathcal{C}_{III}^{2})^{2}k^{2}$ and $\vartheta_{8} = \mathcal{C}_{III}^{1} \mathcal{C}_{III}^{2}(k^{1}+k^{2})/[(\mathcal{C}_{III}^{1})^{2}k^{1} + (\mathcal{C}_{III}^{2})^{2}k^{2}]$.    
In the same way, starting from Eq. (\ref{eq:55}), we obtain
\begin{equation}\label{eq:58}
    \frac{dx}{dt} = \iota_{0} \frac{ \iota_{l} \cosh^{l}{(2 \sigma r_{s})}}{ 1 + \iota_{j}^{\prime} \cosh^{j}{(2 \sigma r_{s})}},
\end{equation}
with $l=1,2,3,4,5$, $j=1,2,3,4,5,6$, $\iota_{0} = [2 \alpha_{0} \beta_{0} (\mu_{1} - \mu_{0})]/[\sqrt{(\beta_{3})^{2} - (\beta_{2})^{2}}]$, $\iota_{1} = 1 - \alpha_{0}$, $\iota_{2} = 2 \alpha_{1}$, $\iota_{3} = 4 \beta_{1} (1 - \alpha_{0}^{2}) + \alpha_{1}^{2}$, $\iota_{4} = 8 \beta_{1} \alpha_{1}$, $\iota_{5} = 4 \beta_{1} \alpha_{1}^{2}$, $\iota_{1}^{\prime} = 2 \alpha_{1}$, $\iota_{2}^{\prime} = 4 (\mu_{0} + \mu_{1}) + \alpha_{1}^{2}$, $\iota_{3}^{\prime} = 8 \alpha_{1} (\mu_{0} + \mu_{1})$, $\iota_{4}^{\prime} = 4 [4 \mu_{0} \mu_{1} + \alpha_{1}^{2} (\mu_{0} + \mu_{1})]$, $\iota_{5}^{\prime} = 32 \alpha_{1} \mu_{0} \mu_{1}$ and $\iota_{6}^{\prime} = 16 \alpha^{2} \mu_{0} \mu_{1}$. However, it is important to highlight that the above differential equation is obtained by considering $0 \leq \mbox{sech}(2 \sigma r_{s})/2 \leq 1/2 \;\; \rightarrow \;\; \cos{[\mbox{sech}(2 \sigma r_{s})/2]} \approx 1$ and $\sin{[\mbox{sech}(2 \sigma r_{s})/2]} \approx \mbox{sech}(2 \sigma r_{s})/2$ in Eq. (\ref{eq:55}). This holds true for $\sigma \ll 1$ \cite{Alcubierre199400}.

But what is the physical or mathematical reason behind this condition? The answer to this question arises when we understand that in this context we are trying to compare the value of $\sigma$ with respect to the radius of the warp bubble, that is, the dimension of the disturbance produced in space-time. However, since this disturbance is generated to transport particles at hyperluminal speeds, when thinking about an electron, the situation at first glance becomes complicated to interpret. Since in orthodox quantum theory the electron exhibits both corpuscular and wave-like behavior, described by the wave function, which does not express the physical size of the electron, but rather the probability of finding it in different regions of space.

Furthermore, the idea of ``size'' for an electron is limited by the probabilistic nature of quantum mechanics and by Heisenberg's uncertainty principle. However, in some contexts, it is possible to speak of an ``effective dimension'' of the electron, related to the spatial extension of its wave function, for example: in an atom, the electron occupies an orbital with a certain shape and size. However, at the fundamental level of Bohmian mechanics, the electron is considered a point particle without internal structure (classical idea), whose dynamics are governed by the guiding wave. Therefore, to give a physical meaning to this last mathematical condition, we consider, without loss of generality, particles of small dimensions, such as an electron, whose classical radius has the value $R_{e} = 2.8178402894(58) \times 10^{-13} \, \mbox{m}$. Comparing it with the Bohr radius ($a_{0} = 5.29177210903(86) \times 10^{-11} \, \mbox{m}$), we obtain the relation $a_{0} = 188 R_{e}$.

\begin{figure}[ht]
    \centering
    \includegraphics[scale=0.7]{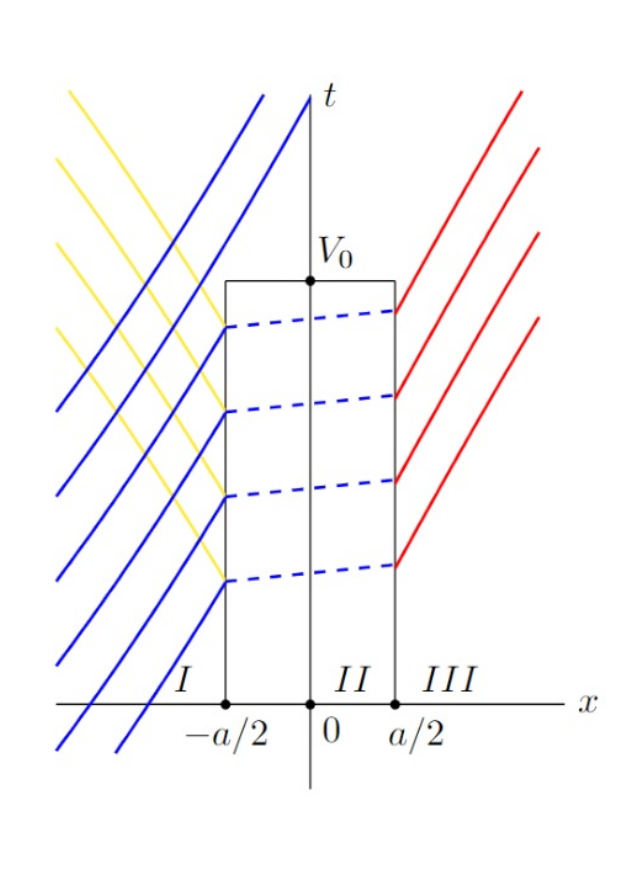}
    \caption{Graphical representation of the trajectories followed by particles incident from the left on the potential barrier.}
    \label{fig:Trayectorias}
\end{figure}

In this sense, when analyzing the expression $\alpha_{1} = 1/\cosh{(2 \sigma R)} = \mbox{sech}(2 \sigma R)$ introduced in Section \ref{sec:4}, from which we derive $\sigma = \mbox{arsech}(\alpha_{1})/2R$ and consequently $\mbox{arsech}(\alpha_{1}) \ll 2R$. It is worth mentioning that, since the quantity $\mbox{arsech}(\alpha_{1})$ is a dimensionless constant, the expression above does not consider the units of $R$ or, in this particular case, $R_{e}$, that is, we only aim to establish a comparison in orders of magnitude. Thus, it is enough to choose a $\alpha_{1}$ such that $R_{e} < \mbox{arsech}(\alpha_{1}) \ll 376 R_{e}$ to obtain an appropriate parameter $\sigma$. Therefore, by solving the differential equations (\ref{eq:56}), (\ref{eq:57}), and (\ref{eq:58}) (see the Appendix), we obtain the expressions
\begin{eqnarray}
    \varrho_{I} &=& \varsigma_{I}^{0} \Xi_{I} + \varsigma_{I}^{1} \tan^{-1}{\biggl[ \varsigma_{I}^{2} \tanh{\left( \frac{\Xi_{I}}{2} \right)} \biggr]}, \label{eq:59} \\
    \varrho_{II} &=& \varsigma_{II}^{0} r_{s} + \frac{2 \sigma^{2}}{3} \varsigma_{II}^{1} r_{s}^{3}, \label{eq:60} \\
    \varrho_{III} &=& \varsigma_{III}^{0} \Xi_{III} + \varsigma_{III}^{1} \tan^{-1}{\biggl[ \varsigma_{III}^{2} \tanh{\left( \frac{\Xi_{III}}{2} \right)} \biggr]}. \label{eq:61}
\end{eqnarray}
Where $\varrho_{I}$, $\varrho_{II}$, $\varrho_{III}$ are arbitrary constants, the coefficients in equations (\ref{eq:59})-(\ref{eq:61}) are defined as follows:
\begin{eqnarray*}
    \varsigma_{I}^{0} &=& \frac{\vartheta_{10}[1 - \vartheta_{11}]}{1 - \vartheta_{12}}, \\
    \varsigma_{I}^{1} &=& \frac{2\vartheta_{10}[\vartheta_{11} - \vartheta_{12}]}{\sqrt{\vartheta_{12}}[1 - \vartheta_{12}]}, \\
    \varsigma_{I}^{2} &=& \frac{1}{\sqrt{\vartheta_{12}}}, \\
    \varsigma_{III}^{0} &=& \frac{\vartheta_{10}^{\prime}[1 - \vartheta_{11}^{\prime}]}{1 - \vartheta_{12}^{\prime}}, \\
    \varsigma_{III}^{1} &=& \frac{2\vartheta_{10}^{\prime}[\vartheta_{11}^{\prime} - \vartheta_{12}^{\prime}]}{\sqrt{\vartheta_{12}^{\prime}}[1 - \vartheta_{12}^{\prime}]}, \\
    \varsigma_{III}^{2} &=& \frac{1}{\sqrt{\vartheta_{12}^{\prime}}}, \\
    \vartheta_{3}^{\prime} &=& \frac{\mathcal{B} \vartheta_{3}}{\mathcal{A}}, \\
    \vartheta_{10} &=& \frac{\vartheta_{3}^{\prime}[1 - \vartheta_{4}][\vartheta_{3}^{\prime} - \vartheta_{1}]}{\vartheta_{3}^{\prime}[1 - \vartheta_{4}] - \vartheta_{1}[1 - \vartheta_{2}]}, \\
    \vartheta_{11} &=& \frac{1 + \vartheta_{4}}{1 - \vartheta_{4}}, \\
    \vartheta_{12} &=& \frac{\vartheta_{3}^{\prime}[1 + \vartheta_{4}] - \vartheta_{1}[1 + \vartheta_{2}]}{\vartheta_{3}^{\prime}[1 - \vartheta_{4}] - \vartheta_{1}[1 - \vartheta_{2}]}, \\
    \vartheta_{10}^{\prime} &=& \frac{\vartheta_{7}[1 - \vartheta_{8}][\vartheta_{7} - \vartheta_{5}]}{\vartheta_{7}[1 - \vartheta_{8}] - \vartheta_{5}[1 - \vartheta_{6}]}, \\
    \vartheta_{11}^{\prime} &=& \frac{1 + \vartheta_{8}}{1 - \vartheta_{8}}, \\
    \vartheta_{12}^{\prime} &=& \frac{\vartheta_{7}[1 + \vartheta_{8}] - \vartheta_{5}[1 + \vartheta_{6}]}{\vartheta_{7}[1 - \vartheta_{8}] - \vartheta_{5}[1 - \vartheta_{6}]}, \\
    \varsigma_{II}^{0} &=& -\frac{1 + \sum_{j=1}^{6} \iota_{j}^{\prime}}{\sum_{j=1}^{6} \iota_{j}^{\prime \prime} - 1}, \\
    \varsigma_{II}^{1} &=& -\frac{\mathcal{N}}{\mathcal{D}^2},
\end{eqnarray*}
\begin{eqnarray*}
    \mathcal{N} &=& (1 + \sum_{j=1}^{6} \iota_{j}^{\prime})(6\iota_{6}^{\prime \prime} - \sum_{j=1}^{6} j \iota_{j}^{\prime \prime}) + (\sum_{j=1}^{6} \iota_{j}^{\prime \prime} - 1)(\sum_{j=1}^{6} j \iota_{j}^{\prime}), \\
    \mathcal{D} &=& \sum_{j=1}^{6} \iota_{j}^{\prime \prime} - 1, \\
    \iota_{j}^{\prime \prime} &=& \iota_{0} \iota_{j} - \iota_{j}^{\prime} \;\; \mbox{(for $j=1,\dots,5$)}, \\
    \iota_{6}^{\prime \prime} &=& \iota_{6}^{\prime}.
\end{eqnarray*}
Then, as the expressions presented in equations (58)-(60) do not allow for solving one variable in terms of the other to obtain the general equation of trajectories for all time $t$, it is possible, at least, to provide an idea of such trajectories for the stationary case by plotting a set of isochronous curves. These curves are obtained by recalling that $\vartheta_{3}^{\prime} = \mathcal{B} \vartheta_{3}/\mathcal{A} = [A - B] \vartheta_{3}/[A + B]$, where $A$ and $B$ are arbitrary constants. This allows rewriting Eq. (58) as the sum of two terms, one positive corresponding to incident particles and one negative corresponding to reflected particles. Thus, the graphical representation of the expressions (58)-(60) is obtained, as shown in Fig. \ref{fig:Trayectorias}.

In Fig. \ref{fig:Trayectorias}, the particle trajectories in the three regimes are presented:
(i) region I (incident in blue and reflected in yellow),
(ii) region II (tunneling, dashed blue line), and
(iii) region III (transmitted). The curves correspond to the analytical solutions of Equations (\ref{eq:58})-(\ref{eq:60}), which for the considered cases reduce to:
$f^{(in)}(x,t) = (1.21x-1)/2 + \tan^{-1}{[10\tan{(1.21x-1)}]}/10 + \varrho_{I}^{(in)}$, $f^{(re)}(x,t) = -(1.21x-1)/2 - \tan^{-1}{[10\tan{(1.21x-1)}]}/10 + \varrho_{I}^{(re)}$, $f^{(tu)}(x,t) = 0.0995(x-2) + \varrho_{II}^{(tu)}$, $f^{(tr)}(x,t) = (1.21x-1)/2 + \tan^{-1}{[10\tan{(1.21x-1)}]}/10 + \varrho_{III}^{(tr)}$.

We emphasize that the position $x_s(t) = 2$, as well as the parameters $\mathcal{B}\Delta k/\mathcal{A}$ and $\Delta E$ in $\Xi_{I}$ and their analogues in $\Xi_{III}$, were chosen for convenience for the graphical representation. The constants $\varsigma_{j}^{\ell_{(\alpha)}}$ (with $\alpha \in \{in, re, tu, tr\}$, $\ell_{(\alpha)} \in \{0,1,2\}$ and $j \in \{I,II,III\}$) were determined by imposing continuity conditions at the potential barrier's boundary points.
The parameters $\varrho$ were discretized to represent different initial conditions: $\varrho_{I}^{(in)} \in \{4.65, 5.65,\dots, 9.65\}$, $\varrho_{I}^{(re)} \in \{-1.75, -0.75, 0.25, 1.25\}$, $\varrho_{II}^{(tu)} \in \{1.75, 2.75, 3.75, 4.75\}$ and $\varrho_{III}^{(tr)} \in \{1.3, 2.3, 3.3, 4.3\}$, with the trajectories closest to the $x$-axis corresponding to the smallest values in each set.

On the other hand, the determination of the tunneling time depends on a comprehensive analysis of some of the quantities of interest introduced elsewhere \cite{Alcubierre199400}, such as the four-velocity and the square of the line element. In particular, we begin with the analysis of the four-velocity of the Eulerian observer given by
\begin{equation}\label{eq:62}
    u^{\mu} = (1,-v_{s}f,0,0),
\end{equation}
whose covariant derivative allows us to obtain the expansion of the volume elements \cite{Alcubierre199400} produced by the bubble
\begin{equation}\label{eq:63}
    \theta = u_{;\nu}^{\mu} = - v_{s} (\partial_{x} f),
\end{equation}
and whose component $u^{1}=v$ mathematically expresses the speed with which the `warp' bubble moves in the $ox$ direction.

For this reason, when considering the low distortion condition ($\sigma \ll 1$) in the bubble profile $f(r_{s})$ (defined in Section~\ref{sec:4}), the warp bubble velocity $v$ in region $II$ (Eulerian observer velocity) is approximated by:
\begin{equation}\label{eq:64-0}
    v = - v_{s} f(r_{s}) \approx - v_{s} \left. \left( \frac{\alpha_{0}}{v_{s} [1 + \alpha_{1} \cosh{(2 \sigma r_{s})}]} \right) \right|_{\sigma \ll 1}.
\end{equation}
Moreover, since $\sigma r_{s} \approx 0$ implies $\alpha_{1} = 2$ and $\cosh(2\sigma r_{s}) \approx 1$ from (\ref{eq:64-0}), we obtain:
\begin{equation}\label{eq:64-1}
    v \approx -\frac{\alpha_{0}}{1 + \alpha_{1}} = -\frac{\alpha_{0}}{3}.
\end{equation}
Then, since $\alpha_{0}$ is defined by $\alpha_{0} = -[v_{s} \tanh{(2\sigma R)}]/[2 \tanh{(\sigma R)}]$ (Section~\ref{sec:4}), and for $\sigma \ll 1$, $\tanh{(\sigma R)} \approx \sigma R$, we have:
\begin{equation}\label{eq:64-2}
    \alpha_{0} \approx - \frac{v_{s} (2 \sigma R)}{2 \sigma R} = - v_{s},
\end{equation}
which when substituted into (\ref{eq:64-1}) yields finally:
\begin{equation}\label{eq:64-3}
    v \approx - \frac{(- v_{s})}{3} = \frac{v_{s}}{3}.
\end{equation}
As $v_{s}$ can take arbitrary values, the `warp' bubble is free to move at even superluminal speeds \cite{Alcubierre199400}. However, if the above statement holds true, the following question becomes valid: would we be in a situation where coordinate time is equal to proper time? The answer to this question arises from analyzing the square of the line element
\begin{equation}\label{eq:65}
    ds^{2} = -dt^{2} + [dx - v_{s}fdt]^{2} + dy^{2} + dz^{2},
\end{equation}
and what it implies to consider a particle inside the bubble moving along the curve $x=x_{s}(t) \rightarrow dx = v_{s}dt$ as the bubble moves through spacetime. Therefore, considering $dy=dz=0$ and $f=1$, we get
\begin{equation}\label{eq:66}
    ds^{2} = - dt^{2} + [v_{s}dt - v_{s}dt]^{2} = - dt^{2}.
\end{equation}
This leads to the following implications. First, the fact that $ds^{2} < 0$ implies that the curve representing the trajectory followed by the particle is of a time-like nature. Second, by rewriting Eq. (\ref{eq:66}), considering the definition of proper time, $d\tau^{2} = - ds^{2}$, we obtain
\begin{equation}\label{eq:67}
    d\tau = dt,
\end{equation}
which implies that the metric proposed in Refs. \cite{Alcubierre199400,Alcubierre200500} does not generate temporal dilations between the time measured with respect to the particle and the coordinate time.

For this reason, we consider a particle moving inside the ``warp'' bubble \cite{Krasnikov199800} following the trajectory $x = x^{\prime} + vt = x^{\prime} + v_{s}/3$ and again $\sigma \ll 1 \rightarrow \sigma^{2} \approx 0$. Consequently, Eq. (\ref{eq:60}) can be rewritten as:
\begin{equation}\label{eq:68}
    \varrho_{II} = \varsigma_{II}^{0} \biggl(x^{\prime} - \frac{v_{s}}{3}t \biggr).
\end{equation}
Then, by evaluating Eq. (\ref{eq:68}) at the entry points $\mathcal{P}_{0}(-a/2,t_{0})$ and exit points $\mathcal{P}_{1}(a/2,t_{1})$, we obtain:
\begin{equation}\label{eq:69}
    \varrho_{II} = \varsigma_{II}^{0} \biggl(- \frac{a}{2} - \frac{v_{s}}{3}t_{0} \biggr)
\end{equation}
and
\begin{equation}\label{eq:70}
    \varrho_{II} = \varsigma_{II}^{0} \biggl(\frac{a}{2} - \frac{v_{s}}{3}t_{1} \biggr),
\end{equation}
which, when subtracted, results in:
\begin{equation}\label{eq:73}
    \Delta t = \frac{3}{v_{s}} a.
\end{equation}
This tunneling time 
depends on the barrier width, as expected; thus, an increase in the potential barrier width shall increase the tunneling time, depending on the value of the tunneling velocity.

In other words, a simple choice like $v_{s} = nc =n$ (since in natural units $c=1$) in (\ref{eq:73}) derived from the model implies that there is no apparent saturation in tunneling times as reported in the literature \cite{Hartman196200,Winful200400,Winful200600,Lantigua202300}. However, it is essential to highlight that the solution (Equation \ref{eq:48}) reproduces the Hartman effect through a different mechanism that must be carefully explained from the obtained results.

To achieve this, an asymptotic analysis of the function
\begin{equation}\label{eq:74}
	f(r_{s}) = \frac{\alpha_{0}}{v_{s}[1 + \alpha_{1} \cosh{(2 \sigma r_{s})}]},
\end{equation}
must be performed, based on the hypothesis of small perturbations ($\sigma \ll 1$), where $r_{s} = x - x_{s}$, $\alpha_{0} \approx - v_{s}$ and $\alpha_{1} \approx 1$. Then, we firstly focus our attention on the inner region of the perturbation, $|r_{s}|<R$. For the regime of small perturbations, the function $f(r_{s})$ admits the following approximation
\begin{equation}\label{eq:75}
	f(r_{s}) \approx \frac{\alpha_{0}}{v_{s}[1 + \alpha_{1}]}
\biggl\{1 - \frac{\alpha_{1}(2 \sigma r_{s})^{2}}{2[1 + \alpha_{1}]} \biggr\},
\end{equation}
where the effective bubble radius $R = [\sqrt{1 + \alpha_{1}}]/[2 \sigma \sqrt{\alpha_{1}}]$ is defined by the criterion $f(R) \approx f(0)/2$, which describes the smooth decay of the spatial distortion. This consideration is analogous to that contemplated by Alcubierre \cite{Alcubierre199400}, where $R$ is determined by the point at which $f(r_{s})$ reaches a significant fraction of its maximum value (not explicitly 1/2, but analogous). This condition, although arbitrary, marks the limit where the spacetime perturbation becomes negligible, ensuring a clear physical transition between the distorted and flat regions.

In the second case, we consider the region external to the bubble, $|r_{s}| \gg R \rightarrow \cosh{(2 \sigma r_{s})} \approx \exp{(2 \sigma r_{s})}/2$, which in this regime of small perturbations the function $f(r_{s})$ admits the following approximation
\begin{equation}\label{eq:76}
	f(r_{s}) \approx \frac{2 \alpha_{0}}{v_{s}\alpha_{1}} \exp{[-(2 \sigma r_{s})]}.
\end{equation}
On the other hand, analyzing the quantum potential $Q_{II}$ (Eq. \ref{eq:49}), we verify that it acts as an effective energy density ($[Q_{II}] = \mbox{Energy}/\mbox{Length}$ in natural units), unlike classical Bohmian mechanics. Furthermore, in the context of the Alcubierre metric, the predominant term in Eq. (\ref{eq:49}) in this regime of small deformations is given by:
\begin{equation}\label{eq:77}
    Q_{II} \approx \frac{v_s^2}{2}f^2(r_s),
\end{equation}
where $f(r_s)$ describes the bubble profile. This formulation implies that:
\begin{itemize}
    \item $Q_{II}$ generates a spacetime distortion whose total energy,
    \begin{equation}\label{eq:78}
	   \mathcal{E} \approx \frac{v_{s}^{2}}{2} \int_{-a/2}^{a/2} dx f^{2}(r_{s}),
    \end{equation}
    scales with the barrier width $a$;
    \item For $a \gg R$, the increase in distortion volume raises $\mathcal{E}$, which in turn amplifies $v_s$;
    \item There is a geometric coupling that ensures $v_s \propto a$, keeping $\Delta t$ constant (Hartman effect).
\end{itemize}

Consequently, $Q_{II}$ not only guides trajectories, but becomes an active source of spacetime deformation.
Then, for narrow barriers $a \leq R$, in this regime of small perturbations ($\sigma \ll 1$), we obtain
\begin{eqnarray*}
    \mathcal{E}_{\leq} &\approx& \frac{v_{s}^{2}}{2} \int_{-a/2}^{a/2} dx  \frac{\alpha_{0}^{2}}{v_{s}^{2}[1 + \alpha_{1}]^{2}} \biggl\{1 - \frac{\alpha_{1}(\cancelto{\ll 1}{2 \sigma r_{s}})^{2}}{2[1 + \alpha_{1}]} \biggr\}^{2} \\
    &=& \frac{\alpha_{0}^{2}}{2[1 + \alpha_{1}]^{2}} \int_{-a/2}^{a/2} dx \\
    &=& \frac{\alpha_{0}^{2}a}{2[1 + \alpha_{1}]^{2}},
\end{eqnarray*}
which, upon substituting $\alpha_{0} \approx - v_{s}$ and $\alpha_{1} \approx 1$, reduces to
\begin{equation}\label{eq:79}
	\mathcal{E}_{\leq} \approx \frac{v_{s}^{2}a}{8}.
\end{equation}
Similarly, for wide barriers $a > 2R$, in this regime of small perturbations, we obtain
\begin{eqnarray}
	\mathcal{E}_{>} &\approx& \frac{v_{s}^{2}}{2} \biggl\{ \frac{4 \alpha_{0}^{2}}{v_{s}^{2} \alpha_{1}^{2}} \exp{(4 \sigma x_{s})} \int_{-a/2}^{-R} dx \exp{(-4 \sigma x)} \nonumber \\
    &+& \int_{-R}^{R} dx f^{2}(0) + \frac{4 \alpha_{0}^{2}}{v_{s}^{2} \alpha_{1}^{2}} \exp{(4 \sigma x_{s})} \int_{R}^{a/2} dx \exp{(-4 \sigma x)} \biggr\} \\
    &=& \frac{v_{s}^{2}}{2} \biggl\{ \frac{4 \alpha_{0}^{2}}{v_{s}^{2} \alpha_{1}^{2}} \exp{(4 \sigma x_{s})} \int_{-a/2}^{-R} dx \exp{(-4 \sigma x)} \nonumber \\
    &+& \frac{\alpha_{0}^{2}}{v_{s}[1 + \alpha_{1}]^{2}} \int_{-R}^{R} dx + \frac{4 \alpha_{0}^{2}}{v_{s}^{2} \alpha_{1}^{2}} \exp{(4 \sigma x_{s})} \int_{R}^{a/2} dr_{s} \exp{(-4 \sigma x)} \biggr\}. \nonumber
\end{eqnarray}
Upon integration and some algebraic procedures, this equation reduces to
\begin{eqnarray}
    \mathcal{E}_{>} &\approx& \alpha_{0}^{2} \biggl\{ \frac{R}{[1 + \alpha_{1}]^{2}} + \frac{\exp{(4 \sigma x_{s})} \exp{(-4 \sigma R)}}{\alpha_{1}^{2} \sigma} \nonumber \\
    &-& \frac{\exp{(4 \sigma x_{s})}\exp{(-2 \sigma a)}}{\alpha_{1}^{2} \sigma}\biggr\},
\end{eqnarray}
from which, after considering $\sigma \ll 1 \rightarrow 4 \sigma R \approx 0$, $4 \sigma x_{s} \approx 0$, $a \gg R \rightarrow 2 \sigma a \gg 2 \sigma R$ and substituting $\alpha_{0} \approx - v_{s}$ and $\alpha_{1} \approx 1$, we finally obtain
\begin{equation}\label{eq:80}
	\mathcal{E}_{>} \approx \frac{v_{s}^{2}}{\sigma}.
\end{equation}

\begin{figure}[ht]
    \centering
    \includegraphics[scale=0.70]{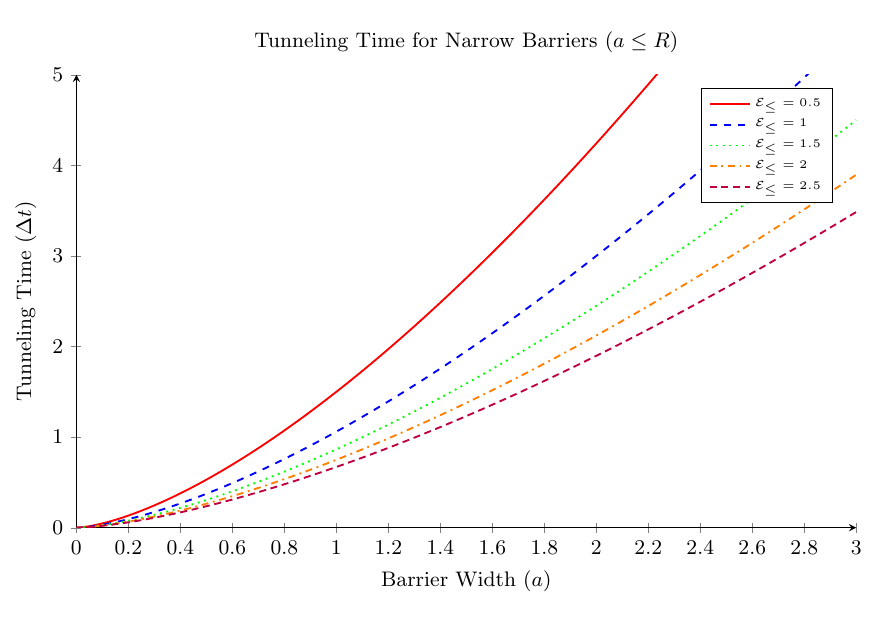}
    \caption{Tunneling time ($\Delta t = 3 [a^{3}/8 \mathcal{E}_{\leq}]^{1/2}$) for narrow barriers ($a \leq R$) with different energy values ($\mathcal{E}_{\leq}$).}
    \label{fig:narrow}
\end{figure}

\begin{figure}[ht]
    \centering
    \includegraphics[scale=0.70]{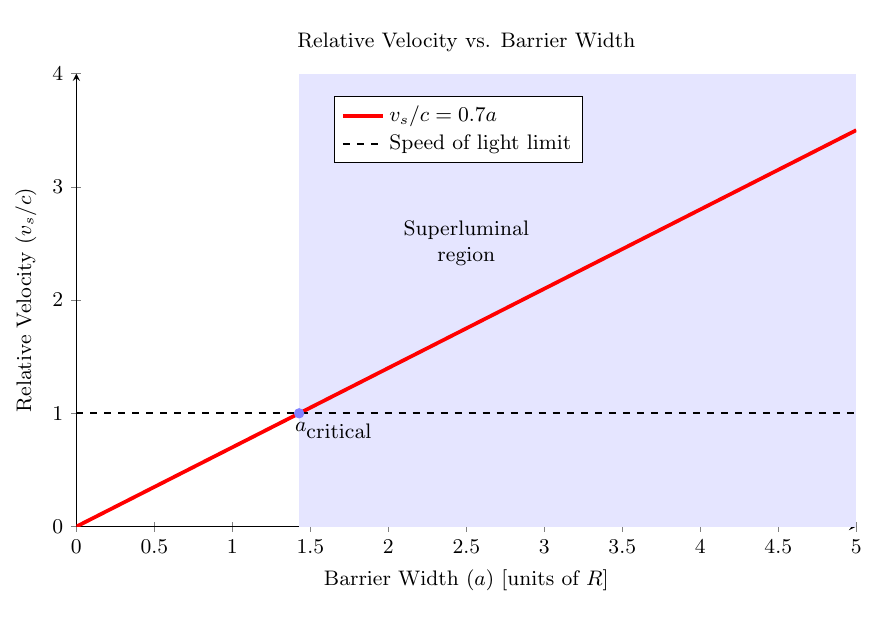}
    \caption{Linear relationship between $v_s/c$ and barrier width $a$. The blue area indicates the superluminal regime ($v_s/c>1$), reached when $a > c/n_0 \approx 1.4R$.}
    \label{fig:velocity}
\end{figure}

Therefore, for narrow barriers, $a \leq R$, from Eq. (\ref{eq:79}) we obtain a velocity $v_{s} \approx \sqrt{8 \mathcal{E}_{\leq}/a}$, which implies a tunneling time $\Delta t = 3 [a^{3}/8 \mathcal{E}_{\leq}]^{1/2}$ (see figure \ref{fig:narrow}), which clearly depends on $a$ (there is no Hartman effect). But for very wide barriers, $a > 2R$, from Eq. (\ref{eq:80}) we obtain a velocity $v_{s} \approx \sqrt{\sigma \mathcal{E}_{>}}$, and since $v_{s} \propto a \rightarrow \sqrt{\sigma \mathcal{E}_{>}} \propto a$. Because of this, the energy must scale as $\mathcal{E}_{>} = [n_{0} a]^{2}/\sigma$ (where $n_{0}$ is a proportionality constant) so that the bubble velocity scales as $v_{s} = n_{0} a$ (see Fig. \ref{fig:velocity}).

\begin{figure}[ht]
    \centering
    \includegraphics[scale=0.70]{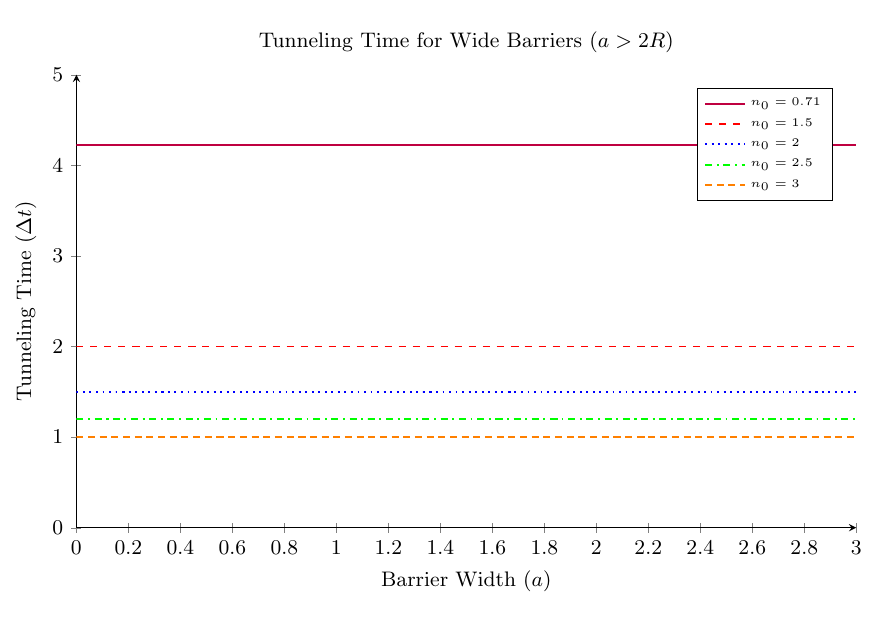}
    \caption{Constant tunneling time ($\Delta t = 3/n_{0}$) (Hartman effect) for wide barriers ($a > 2R$) with different scale factors ($n_{0}$).}
    \label{fig:wide}
\end{figure}

Thus, we obtain the constant tunneling time $\Delta t = 3/n_{0}$ (see Fig. \ref{fig:wide}), in which it is evident that the energy-width coupling implies that as the barrier widens, the quantum potential provides exactly the additional distortion energy needed to maintain a fixed tunneling time. The Alcubierre-type spacetime,  Eq. (\ref{eq:34}), mediates this process through the geometric relationship between $Q_{II}$ and $v_{s}$ in Eq. (\ref{eq:49}), achieving superluminal tunneling (even for low-energy incident particles $E \ll V_{0}$) without violating relativistic principles  \cite{Gron200700,Alcubierre199400}. This explains the saturation of tunneling times (Hartman effect) observed in Refs. \cite{Hartman196200,Winful200400}, through a geometric compensation mechanism.

\section{Concluding remarks}
\label{sec:7}
This work presents a geometrodynamic extension of Bohmian mechanics to explain quantum tunneling, where particles traverse barriers along geodesics in an Alcubierre-type spacetime. The derived model reveals a dual behavior in tunneling dynamics: for narrow barriers, the tunneling time depends on the barrier width, while for wide barriers, it saturates to a constant value—reproducing the Hartman effect. This saturation arises from a geometric self-regulation mechanism, where the quantum potential dynamically adjusts spacetime distortion to maintain a fixed tunneling time, consistent with relativistic causality despite effective superluminal propagation.

The framework bridges quantum mechanics and general relativity by attributing superluminal tunneling to local metric distortions rather than non-local effects or approximations. Crucially, it provides a new theoretical foundation to address longstanding paradoxes, particularly in light of recent experimental evidence, 
which challenges traditional Bohmian predictions (e.g., infinite dwell times in reflective barriers). By linking the Hartman effect to the dynamics of Alcubierre-like spacetime, our model resolves the saturation paradox without ad hoc assumptions, while preserving the causal structure of relativity.

Future directions for this research could explore experimental validation of the model’s predictions in systems where spacetime curvature and quantum tunneling naturally interplay—such as optomechanical cavities or graphene heterostructures—while also investigating connections to geometric phases like the Berry phase in topological materials, where curvature-induced effects may mimic the proposed geometrodynamic tunneling. Although the model relies on negative energy densities (analogous to the Casimir effect), recent advances in metamaterials and quantum field theory, including squeezed vacuum states and hyperbolic metamaterials, suggest promising avenues to engineer these conditions in laboratory settings.

This work not only advances the understanding of spacetime’s role in quantum phenomena but also underscores the need for beyond-standard interpretations to reconcile theory with emerging experimental frontiers. The geometrodynamic approach, though demanding in its mathematical and physical prerequisites, offers a viable pathway to unify quantum non-locality and relativistic constraints, paving the way for tests of quantum-gravity analogies in controlled settings.

\section*{Acknowledgments}
My dearest, first and foremost, I, SL, thank God the Father, infinite goodness, God the Son, redeemer of the world, and God the Holy Spirit, together with the Virgin Mary and Saint Joseph, for their protection, inspiration, and strength on this journey in pursuit of knowledge. I also thank my family, whose unwavering support shaped the person I am today.

This work was made possible through funding from the National Council for Scientific and Technological Development (CNPq), including the grant from the Institutional Training Program (PCI), Process No. 301066/2025-6, awarded by the Ministry of Science, Technology and Innovation (MCTI). Additional support came from the Coordination for the Improvement of Higher Education Personnel (CAPES) under Grant No. 88887.630121/2021-00, as well as CNPq Grants No. 309862/2021-3, No. 409673/2022-6, and No. 421792/2022-1. We also acknowledge the National Institute for the Science and Technology of Quantum Information (INCT-IQ) under Grant No. 465469/2014-0.

To all those who contributed directly or indirectly, we express our deepest gratitude.
\appendix
\section{Solution of the differential equations (\ref{eq:38}), (\ref{eq:56}), (\ref{eq:57}), and (\ref{eq:58})}
\label{sec:8}
In this appendix, the solution to the differential equations (\ref{eq:38}), (\ref{eq:56}), (\ref{eq:57}), and (\ref{eq:58}) is presented. This includes changes of variables, some algebraic procedures, and methods used for the resolution of each of these differential equations \cite{Makarenko198400, Piskunov197700, Piskunov197701}. These expressions have a structure that seemingly leads to complicated differential equations. For this reason, and with the intention of facilitating the reading of this work, we divide this appendix into three parts. The first part focuses on the solution of Eq. (\ref{eq:38}), the second part deals with the solution of Eqs. (\ref{eq:56}) and (\ref{eq:57}), while the third part concentrates on the solution of Eq. (\ref{eq:58}).

\subsection{Solution of Eq. (\ref{eq:38})}
\label{Ap:ap1}

The first step towards solving Eq. (\ref{eq:38}) involves substituting the function $f(r_{s})$ to obtain
\begin{equation}\label{eq:A.1}
    dS(r_{s}) = - 2 \alpha_{0} dr_{s} \frac{1+\alpha_{1} \cosh{(2 \sigma r_{s})}}{ [1+\alpha_{1} \cosh{(2 \sigma r_{s})}]^{2} - \alpha_{0}^{2}}.
\end{equation}
Subsequently, to simplify Eq. (\ref{eq:A.1}) into a more integrable expression, three substitutions were made, which are presented below. First, the substitution $\alpha_{0} \xi = 1+\alpha_{1} \cosh{(2 \sigma r_{s})}$ and $dr_{s} = \alpha_{0} d \xi / [2 \sigma \sqrt{(\alpha_{0} \xi - 1)^{2} - \alpha_{1}^{2}}] $ in Eq. (\ref{eq:A.1}) yields
\begin{equation}\label{eq:A.2}
    dS(\xi) = - \frac{\alpha_{0}}{\sigma} \frac{d \xi }{ \sqrt{(\alpha_{0} \xi - 1)^{2} - \alpha_{1}^{2}}} \frac{\xi}{ \xi^{2} - 1}.
\end{equation}
In the second place, the substitution of $\alpha_{1} \sec{(x)} = \alpha_{0} \xi -1$ and $\sec{(x)} \tan{(x)} dx = \alpha_{0} d \xi / \alpha_{1}$ in Eq. (\ref{eq:A.2}) leads to the expression
\begin{equation}\label{eq:A.3}
    dS(x) = \frac{\alpha_{0}}{\sigma} dx \frac{\alpha_{1} + \cos{(x)}}{ (\alpha_{1} + \cos{(x)})^{2} - \alpha_{0}^{2} \cos^{2}{(x)}}.
\end{equation}
In the third place, through the Weierstrass substitution \cite{Piskunov197700}, which considers $dx = 2 dz / [1+z^{2}]$, $\cos{(x)} = [1-z^{2}]/[1+z^{2}]$, and $z = \tan{(x/2)}$, Eq. (\ref{eq:A.3}) can be reduced to
\begin{equation}\label{eq:A.4}
    dS(z) = \frac{\alpha_{0} \beta_{0}}{2 \sigma \sqrt{\beta_{3}^{2} - \beta_{2}^{2}}} dz \biggl\{ \frac{\beta_{1} - \mu_{0}}{z^{2} + \mu_{0}} - \frac{\beta_{1} - \mu_{1}}{z^{2} + \mu_{1}} \biggr\},
\end{equation}
where the constants $\alpha_{0}$, $\beta_{0}$, $\beta_{1}$, $\beta_{2}$, $\beta_{3}$, $\mu_{0}$, and $\mu_{1}$ were presented in Sec. \ref{sec:4} and are obtained by writing $[\alpha_{1} + \cos{(x)}]/[(\alpha_{1} + \cos{(x)})^{2} - \alpha_{0}^{2} \cos^{2}{(x)}]= \beta_{0} [z^{2} + \beta_{1}]/[z^{4} + 2\beta_{3} z^{2} + \beta_{2}^{2}]$ from the Weierstrass substitution. Consequently, by integrating Eq. (\ref{eq:A.4}), we find
\begin{eqnarray}\label{eq:A.5}
    S(z) &=& \frac{\alpha_{0} \beta_{0}}{2 \sigma \sqrt{\beta_{3}^{2} - \beta_{2}^{2}}} \biggl\{ \frac{\beta_{1} - \mu_{0}}{\sqrt{\mu_{0}}} \tan^{-1}{\left[\frac{z}{\sqrt{\mu_{0}}}\right]} \\
    &-& \frac{\beta_{1} - \mu_{1}}{\sqrt{\mu_{1}}} \tan^{-1}{\left[\frac{z}{\sqrt{\mu_{1}}}\right]} \biggr\}, \nonumber
\end{eqnarray}
which finally leads to the phase function in Eq. (\ref{eq:40}) by reversing the substitutions made previously.

\subsection{Solution of Eqs. (\ref{eq:56}) and (\ref{eq:57})}
\label{Ap:ap2}

To achieve the goal of solving equations (\ref{eq:56}) and (\ref{eq:57}), it must be emphasized that, in addition to multiplicative constants, these expressions have the same structure. Therefore, by finding the solution to one of them, the solution to the second one is obtained by choosing the appropriate multiplicative constants. Subsequently, considering $\Xi_{I} = \mathcal{B} \Delta k x/ \mathcal{A} - \Delta E t = X - T$ and $\vartheta_{3}^{\prime} = \mathcal{B} \vartheta_{3} / \mathcal{A}$, it is possible to write Eq. (\ref{eq:56}) as
\begin{equation}\label{eq:A.6}
    0 = \underbrace{\vartheta_{3}^{\prime} \{ 1 + \vartheta_{4} \cos{\Xi_{I}} \}}_{P}dt \underbrace{- \vartheta_{1} \{ 1 + \vartheta_{2} \cos{\Xi_{I}} \}}_{Q}dx,
\end{equation}
which is evidently not exact since calculating $\partial_{T} P = \partial_{x} Q$ leads to the inequality
\begin{equation}\label{eq:A.7}
    \vartheta_{3}^{\prime} \vartheta_{4} \sin{\Xi_{I}} \neq \vartheta_{1} \vartheta_{2} \sin{\Xi_{I}}.
\end{equation}
However, equation (\ref{eq:A.6}) becomes exact through the integrating factor $\mu(\Xi)$ \cite{Makarenko198400, Piskunov197701}, which is calculated from the equality $\partial_{X}(\mu P) = \partial_{T} (\mu Q)$ and considering $\partial_{X} \Xi_{I} = - \partial_{T} \Xi_{I} = 1$, allowing us to obtain
\begin{equation}\label{eq:A.8}
    \mu(\Xi_{I}) = \frac{1}{1 + \vartheta_{9} \cos{\Xi_{I}}},
\end{equation}
where $\vartheta_{9} = (\vartheta_{3}^{\prime} \vartheta_{4} - \vartheta_{1} \vartheta_{2})/(\vartheta_{3}^{\prime} - \vartheta_{1}) $. Therefore, the problem is reduced to integrating the expression
\begin{equation}\label{eq:A.9}
    d U(\Xi_{I}) = d \Xi_{I} \frac{ \vartheta_{3}^{\prime} + \vartheta_{3}^{\prime} \vartheta_{4} \cos{\Xi_{I}}}{1 + \vartheta_{9} \cos{\Xi_{I}}},
\end{equation}
which, through the Weierstrass substitution  \cite{Piskunov197700}, considering $d\Xi_{I} = 2 dz / [1+z^{2}]$, $\cos{\Xi_{I}} = [1-z^{2}]/[1+z^{2}]$, and $z = \tan{(\Xi_{I}/2)}$, can be reduced to
\begin{equation}\label{eq:A.10}
    d U(z) = \frac{ 2 \vartheta_{10}}{1 - \vartheta_{12}} \biggl\{ [1 - \vartheta_{11}] \frac{dz}{1 + z^{2}} + [\vartheta_{11} - \vartheta_{12}] \frac{dz}{\vartheta_{12} + z^{2}} \biggr\},
\end{equation}
where $\vartheta_{10} = \vartheta_{3}^{\prime}[1 - \vartheta_{4}]/[1 - \vartheta_{9}]$, $\vartheta_{11} = [1 + \vartheta_{4}]/[1 - \vartheta_{4}]$, and $\vartheta_{12} = [1 + \vartheta_{9}]/[1 - \vartheta_{9}]$.

Then, by integrating Eq. (\ref{eq:A.10}), we obtain
\begin{eqnarray} \label{eq:A.11}
    U(\Xi_{I}) &=& \frac{ 2 \vartheta_{10}}{1 - \vartheta_{12}} \biggl\{ \frac{[1 - \vartheta_{11}]}{2} \Xi_{I} + \frac{[\vartheta_{11} - \vartheta_{12}]}{\sqrt{\vartheta_{12}}} \\
    &\times& \tan^{-1}{ \biggl[ \frac{1}{\sqrt{\vartheta_{12}}} \tan{ \biggl( \frac{\Xi_{I}}{2} \biggr)} \biggr] } \biggr\} + U_{0}, \nonumber
\end{eqnarray}
from which the general solution of the exact version of Eq. (\ref{eq:56}) is obtained by determining the value of $U_{0}(T)$. Therefore, by calculating $\partial_{T} U$ and equating it with the expression $\mu Q$, it follows that $U_{0}(T)=0$, as it is not possible to obtain a function that depends only on $T$ from the previous calculation.
On the other hand, to determine the general solution from Eq. (\ref{eq:57}), it suffices to replace $\vartheta_{1} \rightarrow \vartheta_{5}$, $\vartheta_{2} \rightarrow \vartheta_{6}$, $\vartheta_{3}^{\prime} \rightarrow \vartheta_{7}$, $\vartheta_{4} \rightarrow \vartheta_{8}$, and $\Xi_{I} \rightarrow \Xi_{III}$ in each of the constants and in the angle of the general solution (\ref{eq:A.11}).

\subsection{solution of Eq. (\ref{eq:58})}
\label{Ap:ap3}

In a manner similar to that used in \ref{Ap:ap2}, equation (\ref{eq:58}) is reformulated as follows:
\begin{equation}\label{eq:A.12}
    0 = \underbrace{  \{\iota_{j}^{\prime} \cosh^{j}{(r)} + 1\} }_{P} dx \underbrace{ - \iota_{0} \iota_{l} \cosh^{l}{(r)} }_{Q} dt,
\end{equation}
with $j=1,2,3,4,5,6$ and $l=1,2,3,4,5$. Then, following a procedure similar to that applied in \ref{Ap:ap2}, we consider $2 \sigma r_{s} = 2 \sigma (x - x_{s}(t)) = X - T \equiv r$, allowing us to calculate the partial derivatives:
\begin{equation}\label{eq:A.13}
    \partial_{T}P(r) = - j \iota_{j}^{\prime} \cosh^{j-1}{(r)} \sinh{(r)}
\end{equation}
and
\begin{equation}\label{eq:A.14}
    \partial_{X}Q(r) = - l\iota_{0} \iota_{l} \cosh^{l-1}{(r)} \sinh{(r)},
\end{equation}
clearly indicating that the differential equation (\ref{eq:A.12}) is not exact. Therefore, after calculating the integration factor \cite{Makarenko198400, Piskunov197701}
\begin{equation}\label{eq:A.15}
    \mu(r) = \{\iota_{j}^{\prime \prime} \cosh^{j}{(r)} - 1\}^{-1},
\end{equation}
Eq. (\ref{eq:A.12}) transforms into an exact equation, where
\begin{equation}\label{eq:A.16}
    \mu(r) P(r) = \frac{\iota_{j}^{\prime} \cosh^{j}{(r)} + 1}{\iota_{j}^{\prime \prime} \cosh^{j}{(r)} - 1}
\end{equation}
and
\begin{equation}\label{eq:A.17}
    \mu(r) Q(r) = - \frac{\iota_{0} \iota_{l} \cosh^{l}{(r)}}{\iota_{j}^{\prime \prime} \cosh^{j}{(r)} - 1},
\end{equation}
meaning that the problem is once again reduced to integrating the expression
\begin{equation}\label{eq:A.18}
    dU(r) = - dr \{ 2 \sigma \}^{-1} \frac{\iota_{j}^{\prime} \cosh^{j}{(r)} + 1}{\iota_{j}^{\prime \prime} \cosh^{j}{(r)} - 1}.
\end{equation}
The exact primitive of (\ref{eq:A.18}) is analytically challenging, but a systematic approximation can be developed for the small-parameter regime ($\sigma \ll 1$). Recalling that $r \equiv r(X,T) = 2 \sigma r_{s}(x,t)$ depends on both spatial and temporal coordinates, we perform a second-order bivariate Taylor expansion \cite{Piskunov197701} of $\mu(r)P(r) = - [\iota_{j}^{\prime} \cosh^{j}{(r)} + 1]/[\iota_{j}^{\prime \prime} \cosh^{j}{(r)} - 1]$ around $\mathcal{P}^{\prime}(0,0)$:
\begin{eqnarray*}
    &&\mu(r)P(r) = \underbrace{\left. \mu P \right|_{(0,0)}}_{\varsigma^{0}} + \underbrace{\left. \frac{\partial (\mu P)}{\partial X} \right|_{(0,0)} \!\!\! X + \left. \frac{\partial (\mu P)}{\partial T} \right|_{(0,0)} \!\!\! T}_{\text{1st-order terms}} \\
    &+& \underbrace{\frac{1}{2} \left. \frac{\partial^2 (\mu P)}{\partial X^2} \right|_{(0,0)} \!\!\! X^2 + \left. \frac{\partial^2 (\mu P)}{\partial X \partial T} \right|_{(0,0)} \!\!\! XT + \frac{1}{2} \left. \frac{\partial^2 (\mu P)}{\partial T^2} \right|_{(0,0)} \!\!\! T^2}_{\text{2nd-order terms}} \\
    &+& \mathcal{O}(X^3, X^2T, XT^2, T^3),
\end{eqnarray*}
where:
\begin{itemize}
    \item The zeroth-order term $\varsigma^{0} = \left. \mu P \right|_{(0,0)}$ captures the static background,
    \item The first-order derivatives vanish at $\mathcal{P}^{\prime}(0,0)$,
    \item The second-order coefficients govern the main nonlinear correction.
\end{itemize}
Higher-order terms ($\mathcal{O}(X^3,X^2T,XT^2,T^3)$) are negligible under the low perturbation condition $\sigma \ll 1$, reducing the expansion to $\mu(r)P(r) \approx \varsigma^{0} + \varsigma^{1} r^{2}/2$. Substituting this into (\ref{eq:A.18}) and integrating yields to:
\begin{eqnarray}\label{eq:A.20}
    U(r) &=& \{ 2 \sigma \}^{-1} \{ \varsigma^{0} r + \frac{ \varsigma^{1} }{6} r^{3} \} + c(t) \\
    &=& \{ 2 \sigma \}^{-1} \{ \varsigma^{0} r + \frac{ \varsigma^{1} }{6} r^{3} \}. \nonumber
\end{eqnarray}
where the integration constant $c(t) = 0$ is required for consistency with $\partial_{x} U = \mu(r) Q(r)$, as in the analogous procedures shown in \ref{Ap:ap2}.

\end{document}